\documentclass[11pt,english]{article}
\usepackage[T1]{fontenc}
\usepackage[latin9]{inputenc}
\usepackage{geometry}
\usepackage{color}
\usepackage{babel}
\usepackage{verbatim}
\usepackage{url}
\usepackage{amsmath}
\usepackage{amsthm}
\usepackage{amssymb}
\usepackage{graphicx}
\usepackage{setspace}
\usepackage[authoryear]{natbib}
\usepackage{verbatim}
\usepackage{multirow}

\usepackage[unicode=true,pdfusetitle,
 bookmarks=true,bookmarksnumbered=false,bookmarksopen=false,
 breaklinks=false,pdfborder={0 0 0},pdfborderstyle={},backref=false,colorlinks=false]
 {hyperref}
\usepackage{breakurl}

\makeatletter

\providecommand{\tabularnewline}{\\}

\usepackage{appendix}

 \theoremstyle{definition}
\newtheorem{example}{\protect\examplename}

\providecommand{\examplename}{Example}

\makeatother
  
\begin{document}
\title{Structural Estimation of Behavioral Heterogeneity}
\author{Zhentao Shi and Huanhuan Zheng}
\date{}
\maketitle

\thispagestyle{empty}
\begin{center}
\textbf{\large{}Abstract}
\par\end{center}{\large \par}

We develop a behavioral asset pricing model in which agents trade
in a market with information friction. Profit-maximizing agents switch
between trading strategies in response to dynamic market conditions.
Due to noisy private information about the fundamental value, the
agents form different evaluations about heterogeneous strategies.
We exploit a thin set\textemdash a small sub-population\textemdash to
pointly identify this nonlinear model, and estimate the structural
parameters using extended method of moments. Based on the estimated
parameters, the model produces return time series that emulate the
moments of the real data. These results are robust across different
sample periods and estimation methods. 

\vspace{0.8cm}



\vspace{0.8cm}

\small \noindent Zhentao Shi (corresponding author): \texttt{zhentao.shi@cuhk.edu.hk},
Department of Economics, 912 Esther Lee Building, the Chinese University
of Hong Kong, Shatin, New Territories, Hong Kong SAR, China. Tel:
(852) 3943-1432. Fax (852) 2603-5805. Huanhuan Zheng: \texttt{sppzhen@nus.edu.sg},
Lee Kuan Yew School of Public Policy, National University of Singapore,
469C Bukit Timah Road, Singapore 259772. We benefit from in-depth
discussion with Taisuke Otsu. We thank Zhenyu Gao, Oliver Linton,
Peter Phillips,  Michael Zheng Song and Jun Yu for helpful comments.
All remaining errors are ours.

\newpage{}

\normalsize

\section{Introduction}

Financial markets undergo cycles of booms and busts. Price fluctuations
generate profit opportunities for different investment strategies.
No single investment strategy can always triumph\textemdash they also
experience cycles of gain and loss in response to shifting market
environment. It is essential for profit-seeking investors to choose
their strategies according to the dynamic market conditions. We try
to understand, theoretically and empirically, the impact of information
friction on strategy switching. Our model follows the common approach
in heterogeneous agent models (HAM), in which an agent selects from
multiple investment principles such as the fundamental and technical
trading strategies, while we introduce information friction to generate
endogenous switching between different strategies. 

In this model, every agent receives a private signal\textemdash an
unbiased forecast about the fundamental value of the risky asset.
Given the presence of information dispersion embodied in the realization
of the private signal, the agents conceive different evaluations for
the same investment strategy. As a result, each agent chooses, from
a set of investment strategies, the one that maximizes the expected
profit. The agents' actions reshape the asset price, and the evolutionary
environment forces the agents to revise their subsequent choices in
the next period. Such dynamic interaction between the agents and the
asset price induces behavioral heterogeneity among agents and boom-bust
cycles in the financial market.

We formally identify the proposed behavioral model via a \emph{thin
set}\textemdash a small subset of the population that reflects some
special cases of the model \citep{khan2010irregular}. We combine
the unconditional moments, which involve the whole sample, with those
conditional moments motivated from the thin sets. The two kinds of
moments differ in the rates of convergence, so that the standard asymptotic
theory for generalized method of moments (GMM) is not directly applicable.
We employ \emph{extended method of moments} (XMM) \citep{gagliardini2011efficient}
for estimation and statistical inference. 

Applying XMM to historical observations of the Standard and Poor 500
index (S\&P 500), we estimate and test our structural model in several
sample periods. The predicted returns from the model closely match
the real data in terms of the mean, standard deviation, skewness and
kurtosis. Furthermore, we find empirical evidence that supports the
evolutionary trading heterogeneity driven by information dispersion.
When the price is relatively close to the fundamental value, an investment
strategy based on the historical price trend is popular in the market.
When the asset is excessively mispriced, however, the agents tend
to switch to a fundamental strategy to pursue higher expected profits;
their collective actions gradually drive the price toward the fundamental
value, which corrects the market.

Our paper makes several contributions to the literature. In terms
of modeling, the dynamics in trading heterogeneity has been modeled
by latent boom-burst market states \citep{chiarella2012estimating},
real business cycles \citep{lof2012heterogeneity}, and switching
stochastic processes \citep{brock1998heterogeneous}. In particular,
Markov transition of discrete regimes is popular in modeling the switching
processes, and finds many empirical applications in the stock market,
commodity market and derivative market \citep{frijns2010behavioral,jongen2012explaining,ter2013dynamic,eichholtz2015fundamentals}.
While these empirical papers directly model the aggregate time series,
they leave unexplained why some agents switch their strategies but
the others do not. Our new model combines \citet{he2016trading}'s
microeconomic mechanism that endogenizes the Markov switching process
and  \citet{hirshleifer1992managerial}'s prioritization of profit
instead of utility for the agency problem in asset management. Built
on a microeconomic foundation of individual behavior, our theoretical
model provides implication of the aggregate time series.

Econometric identification is the bridge that links the economic structural
model and the data. Well-known is the difficulty to check identification
in nonlinear models \citep{rothenberg1971identification,newey1994large,komunjer2012global}.
Formal identification is largely missing in the literature of HAM,
where nonlinearity is the rule rather than the exception. While following
the convention of HAM, we construct our model with identification
in mind. The switching between the fundamental and technical strategies
opens the opportunity for us to scrutinize in ``slow motion'' the
instant, or the thin set, when the market is overwhelmed by one strategy.
When a single strategy dominates, identification can be easily verified.
We explore the thin-set identification and manage to recover all structural
parameters in our model. To the best of our knowledge, this is the
first paper that formally analyzes and establishes identification
in the literature of structural modeling of heterogeneous behavior
in the financial market.

In terms of estimation methods, existing empirical works of HAM mostly
use nonlinear least squares \citep{boswijk2007behavioral,chiarella2012estimating,frijns2010behavioral},
except that \citet{franke2012structural} utilize the simulated method
of moments (SMM). We derive an explicit formula of the pricing mechanism
that implies moment restrictions in closed-form, which simplifies
and speeds up the estimation. XMM is exactly the right bottle opener
for a champagne brewed by the thin-set identification, thanks to the
econometricians who crafted it. 

The rest of the paper is organized as follows. Section \ref{sec:Information-Based-Structural-Mod}
develops the information-driven structural asset pricing model of
behavioral heterogeneity. Section \ref{sec:empirical} discusses identification,
data handling, and estimation. Section \ref{sec:Empirical-results}
reports the empirical findings, and compares them with those based
on alternative approaches. Section \ref{sec:Conclusion} concludes
the paper. Moreover, we have prepared an Online Supplement with additional
empirical results, extension, implementation, and examples. 

\section{Information-Based Structural Model\label{sec:Information-Based-Structural-Mod}}

In this section, we summarize the key building blocks of the information-based
structural model. Step-by-step derivation of the model is given in
Appendix Section \ref{sec:Complete-Description-of }.  A continuum
(of measure one) of agents trade on one risky asset and one risk-free
asset. The logarithm of the fundamental value of the risky asset at
period $t$, denoted as $\mu_{t}$, is an exogenous random variable
that market participants cannot interfere. It follows a random walk
$\mu_{t}=\mu_{t-1}+\sigma_{\mu}\varepsilon_{t}^{\mu}$ for some $\sigma_{\mu}>0$,
where $\varepsilon_{t}^{\mu}$ is independently and identically distributed
across $t$ with mean and variance standardized as 0 and 1, respectively.
Let $\boldsymbol{\mu}^{t}=\left(\mu_{t},\mu_{t-1},\mu_{t-2},\ldots,\mu_{0}\right)$
be the history of the fundamental value. At the beginning of period
$t$, each agent receives a private signal $x_{it}=\mu_{t}+\sigma_{x}\varepsilon_{it}$,
an unbiased forecast of the fundamental value $\mu_{t}$. The noise
$\varepsilon_{it}|\boldsymbol{\mu}^{t}\sim\mathrm{i.i.d.}\Lambda$,
where $\Lambda$ is a strictly increasing distribution function with
the support of the real line, its density symmetric around 0, and
the variance standardized as 1. 

Let $\mathbf{p}^{t-1}=\left(p_{t-1},p_{t-2},\ldots,p_{0}\right)$
be the logarithm of the past price. Both $\mathbf{p}^{t-1}$ and $\boldsymbol{\mu}^{t-1}$
are public information for all investors at the beginning of time
$t$. Each agent consults two financial advisors who conduct fundamental
analysis and chartist analysis independently, to which we refer as
$f$-advisor and $c$-advisor, respectively. The advisors make forecast
according to their own perception of price movement, which may not
be consistent with the true price formation mechanism. The $f$-advisor
expects the price to respond to the fundamental value. Once she learns
the private information $x_{it}$ from her client, she updates the
expected $\mu_{t}$ to be $\frac{\mu_{t-1}+\alpha x_{it}}{1+\alpha}$,
which is an average of $\mu_{t-1}$ and $x_{it}$ weighted by the
precision (the inverse of variance), where $\alpha=\sigma_{\mu}^{2}/\sigma_{x}^{2}$
measures the precision of private information relative to public information.
Believing in the efficient market hypothesis, she expects the period-$t$
return to be $\frac{\alpha\sigma_{x}}{1+\alpha}\left(\varepsilon_{it}-\delta_{t}\right),$
where $\delta_{t}=\left(\left(1+\alpha\right)p_{t-1}-\mu_{t-1}-\alpha\mu_{t}\right)/\left(\alpha\sigma_{x}\right)$.
The $f$-advisor maximizes the constant absolute risk aversion (CARA)
utility function and recommends the optimal investment flow $q_{it}^{f*}=\eta\frac{\alpha\sigma_{x}}{1+\alpha}\left(\varepsilon_{it}-\delta_{t}\right)$
into the risky asset, where $\eta$ is the trading intensity of the
fundamental strategy with respect to asset mispricing. 

In the meantime, the $c$-advisor utilizes technical analysis to forecast
price movement. Her strategy is based only on the historical price
trend, rather than $x_{it}$ or $\boldsymbol{\mu}^{t-1}$. Her expected
period-$t$ return is $\Delta_{t-1}=p_{t-1}-p_{t-1}^{\mathrm{ref}}$,
where $p_{t-1}^{\mathrm{ref}}$ is the reference price derived from
certain technical rules. Under the same utility function, the $c$-advisor
recommends the optimal investment flow $q_{t}^{c*}=\tau\Delta_{t-1}$
into the risky asset, where $\tau$ is the trading intensity of the
chartist strategy. Unlike $q_{it}^{f*}$ that varies with $i$, for
each individual $q_{t}^{c*}$ is the same.

We focus on the fundamental and technical strategies of bounded rationality
out of many alternatives for the following reasons. (i) The two strategies
are used commonly in practice \citep{allen1990charts}. (ii) Models
accounting for such two strategies are powerful in explaining financial
market phenomena such as bubbles and crashes \citep{lux1995herd,huang2010financial}
and providing empirical specifications that outperform random walk
\citep{chiarella2012estimating}. (iii) Due to resource constraints,
it is reasonable to prioritize investment strategies with good tracking
records, supported by theoretical or empirical foundations; it is
costly to hire a large number of financial advisors to conduct various
analysis. (iv) No evidence suggests that other types of analysis consistently
outperform fundamental and technical analysis in terms of profitability
or utility.

Neither strategy is rational in that they ignore how agents' trading
behavior affects the price. Forming rational expectation is difficult
in the current setup due to the uncertainty about the convergence
of the price to the fundamental value. It deviates from the rational
expectation model, in which the price must return to its value at
the terminal period. The fundamental strategy that utilizes private
information does not always dominate the chartist strategy because
the price\textemdash determined by the aggregate action of market
participants\textemdash may not necessarily reflect the information. 

Next, we discuss how the agents select trading strategies. Unlike
the financial advisors who care about utility, the agents seek to
maximize their investment profit in excess to the risk-free asset
\citep{hirshleifer1992managerial}.\footnote{Allowing the agents to have a different target function from financial
advisors highlights the contrast between practitioners, who adopt
straightforward criteria to swiftly respond to market, and researchers,
who focus on sophisticated measures of utility. Our main results still
hold when agents maximize CARA utility as their financial advisors
do.} Let $\pi_{it}^{f}$ be the expected profit of the fundamental strategy
based on the information available at the beginning of period $t$,
and $\pi_{t}^{c}$ be that of the chartist strategy. In our model,
we have 
\begin{equation}
\begin{array}{l}
\pi_{it}^{f}=\eta\left(\frac{\alpha\sigma_{x}}{1+\alpha}\left(\varepsilon_{it}-\delta_{t}\right)\right)^{2}\\
\pi_{t}^{c}=\tau\Delta_{t-1}^{2}.
\end{array}\label{eq:pi_fc}
\end{equation}

An agent chooses the strategy that yields higher expected profit.
Due to the constraints on risk exposure and resources, we assume that
every agent adopts one and only one strategy. Investors are not confident
to select strategies that they are unfamiliar with, especially those
insufficiently corroborated by studies or experience. It is therefore
reasonable to presume that agents ignore strategies that are not scrutinized
by their financial advisors. 

Now we set equal $\pi_{it}^{f}$ and $\pi_{t}^{c}$ to solve the threshold
signals that make agents indifferent between the fundamental and the
chartist strategy. The quadratic form in (\ref{eq:pi_fc}) yields
the lower bound $\overline{\varepsilon}_{t}^{m}=\delta_{t}-\zeta_{t-1}$
and upper bound $\bar{\varepsilon}_{t}^{M}=\delta_{t}+\zeta_{t-1}$,
where $\zeta_{t-1}=\frac{1+\alpha}{\alpha\sigma_{x}}\sqrt{\frac{\tau}{\eta}}\left|\Delta_{t-1}\right|$.
The individual choice of the strategy hinges on the private signal.
We assume that the agent will choose the fundamental strategy when
she is indifferent between the two options. When $\varepsilon_{it}\in(-\infty,\overline{\varepsilon}_{t}^{m}]\cup[\overline{\varepsilon}_{t}^{M},\infty),$
the agent will adopt the fundamental strategy and we call her a \textit{fundamentalist}.
When $\varepsilon_{it}\in\left(\overline{\varepsilon}_{t}^{m},\overline{\varepsilon}_{t}^{M}\right),$
she will take the chartist strategy, and we call her a \emph{chartist}.
Given the distribution of the private signal, the fraction of chartists
is
\[
m_{t}=\Lambda\left(\overline{\varepsilon}_{t}^{M}\right)-\Lambda\left(\overline{\varepsilon}_{t}^{m}\right).
\]
 The fraction of fundamentalists is $1-m_{t}$. If in addition $\Lambda$
is unimodal, an application of the Leibniz integral rule to $m_{t}$
shows that it strictly decreases in $\left|\delta_{t}\right|\in\left(0,\infty\right)$.
Since $\left|\delta_{t}\right|$ captures the degree of mispricing,
the fraction of chartists is relatively large (small) when the market
is moderately (excessively) mispriced. 

After selecting their preferred strategies at the beginning of period
$t$, all agents place their trading orders simultaneously to a market
maker. Following \citet{lux1995herd}, we assume that the market maker
adjusts the price according to 
\[
p_{t}\left(\theta\right)=p_{t-1}+\rho D_{t}\left(\theta\right),
\]
 where $\rho>0$ is the marginal impact of aggregate demand on the
asset price, $\theta=\left(\eta,\tau,\alpha,\sigma_{\mu}\right)$
is the set of the other structural parameters,\footnote{Since $\sigma_{x}=\sigma_{\mu}/\sqrt{\alpha}$, we do not need to
include $\sigma_{x}$ into $\theta$ given the presence of $\sigma_{\mu}$
and $\alpha$.} and 
\begin{equation}
D_{t}\left(\theta\right)=\frac{\eta\alpha\sigma_{x}}{1+\alpha}\left[\varphi\left(\bar{\varepsilon}_{t}^{m}\right)-\varphi\left(\bar{\varepsilon}_{t}^{M}\right)-\left(1-m_{t}\right)\delta_{t}\right]+\tau m_{t}\Delta_{t-1},\label{eq:demand}
\end{equation}
is the aggregate demand in the market, where $\varphi\left(a\right)=\int_{-\infty}^{a}zd\Lambda\left(z\right)$
is the upper-truncated mean. According to the model, the stock market
return follows 
\begin{equation}
R_{t}\left(\theta\right)=p_{t}\left(\theta\right)-p_{t-1}=\rho D_{t}\left(\theta\right).\label{eq_Return}
\end{equation}
The above equation characterizes the asset price movements. It will
be the key equation for the empirical estimation. 

We apply the market-maker framework, instead of the market-clearing
mechanism, because the former enables the nonlinear model to be analytically
tractable over multiple horizons while the latter does not necessarily
yield a solution for the equilibrium price. \citet{10.2307/27647318}
find that the market-maker mechanism performs as well as, if not outperforms,
the market-clearing mechanism in terms of generating the efficient
price. 

We conclude this section by comparing our model with the Markov regime-switching
regression. The Markov switching model is originated from \citet{hamilton1989new},
and has been extended over the decades \citep{kim1994dynamic,kim1999has},
with the latest development endogenizing the latent state variable
\citep{kim2008estimation,chang2016new}. Regime-switching models are
featured by the transition probability among discrete states. In contrast,
the microeconomic mechanism in our model dictates the variation of
the fraction of agents who adopt either strategy in the dynamic market
environment. On the one hand, our approach preserves the Markov property
since $R_{t}\left(\theta\right)$ depends only on $\left(p_{t-1},p_{t-1}^{\mathrm{ref}},\mu_{t},\mu_{t-1}\right)$,
which the econometrician directly observes when analyzing the data,
but no other past observations. On the other hand, our approach differs
from the Markov regime-switching regression as we do not directly
model the aggregate time series. Instead, the aggregate market demand
is generated by summing up the individual demand. Furthermore, an
agent's switching between heterogeneous strategies is endogenous,
because the threshold of the strategy choice is implied by the profit
maximization problem. In other words, we attempt to provide a microeconomic
foundation for the association between the latent states and the aggregate
time series.

\section{Econometric Methodology\label{sec:empirical}}

A model is judged not only by its microeconomic foundation, but also
by its empirical fitness. We push the model to encounter data in this
section. We verify that the structural parameters can be identified
from the distribution of the observable random variables, and then
propose an estimation procedure.

\subsection{Thin-Set Identification \label{subsec:thin-set-Identification}}

The structural model is a description of the data generating process,
while the analysis of identification bridges the gap between the theoretical
model and the observed data. The unobservable noises in the structural
model stem from $\left(\varepsilon_{t}^{\mu},\varepsilon_{it}\right)$,
which are independently and identically distributed across time. As
a result, $\left(R_{t}\left(\theta\right)=\rho D_{t}\left(\theta\right)\right)_{t=1}^{T}$
is strictly stationary according to the model. 

In reality, the econometrician observes two time series $\mathbf{p}^{T}$
and $\boldsymbol{\mu}^{T}$. If the observable random variables are
truly generated from the theoretical model, can we uniquely determine
the value of the ``deep parameters'' $\left(\sigma_{\mu},\eta,\tau,\alpha,\rho\right)$
from the joint distribution of $\left(\mathbf{p}^{T},\boldsymbol{\mu}^{T}\right)$?
Obviously, $\sigma_{\mu}$ can be directly identified from $\boldsymbol{\mu}^{T}$.
We narrow down the question to recovering the parameters $\left(\eta,\tau,\alpha,\rho\right)$
by matching the distribution of $\left(R_{t}\left(\theta\right)\right)_{t=1}^{T}$,
which comes from the theory, with the distribution of the observable
$\left(R_{t}^{\text{r}}=p_{t}-p_{t-1}\right)_{t=1}^{T}$, where the
superscript ``r'' stands for ``real''. Nevertheless, $\left(\eta,\tau,\rho\right)$
\emph{cannot} be identified jointly. In view of  (\ref{eq:demand})
and (\ref{eq_Return}), if we multiply $\rho$ by a non-zero constant
and divide $\eta$ and $\tau$ by the same constant, the resulting
$R_{t}\left(\theta\right)$ in (\ref{eq_Return}) remains. Hence we
have to normalize $\rho=1$ and discuss the identification of the
other three parameters $\left(\eta,\tau,\alpha\right)$. 

It is well-known that global identification is often difficult in
nonlinear models \citep{rothenberg1971identification,newey1994large,komunjer2012global}.
In the literature of HAM, identification of structural parameters
is largely ignored. In this paper, we formally establish point identification
for this highly nonlinear structural model. We take the \emph{thin-set
identification} approach \citep{khan2010irregular,lewbel2016identification},
conditioning on some events that occur on a set of measure zero if
the random variables are continuously distributed.\footnote{This thin-set identification strategy is not peculiar to our model.
In Supplement Section S5, we provide examples in which thin-set identification
can be invoked to establish point identification for other HAM models. } The key insight for the point identification is that when the event
\[
G_{1}=\left\{ \Delta_{t-1}=0\right\} 
\]
 occurs, the expected return of the chartist strategy becomes zero,
and all investors thereby turn to the fundamental strategy. Conditional
on $G_{1}$, we have $\bar{\varepsilon}_{t}^{m}=\bar{\varepsilon}_{t}^{M}$
and $m_{t}=0$, and can simplify  (\ref{eq_Return}) as 
\begin{equation}
R_{t}\left(\theta\right)=\theta_{1}\tilde{z}_{1,t}+\theta_{2}\tilde{z}_{2,t},\label{eq:fun_only}
\end{equation}
where $\tilde{z}_{1t}=\mu_{t-1}-p_{t-1}$ and $\tilde{z}_{2t}=\mu_{t}-p_{t-1}$
are observable, and $\theta_{1}=\eta/\left(1+\alpha\right)$ and $\theta_{2}=\eta\alpha/\left(1+\alpha\right)$
are explicit functions of the deep parameters. As long as the conditional
distribution $\left(\tilde{z}_{1t},\tilde{z}_{2t}\right)\big|G_{1}$
is not perfectly collinear, we can identify $\theta_{1}$ and $\theta_{2}$,
and then recover $\alpha=\theta_{2}/\theta_{1}$ and $\eta=\theta_{1}+\theta_{2}$.
The occurrence of $G_{1}$ highlights the particular instant when
the market is overwhelmed by the fundamental strategy, and the identification
of $\alpha$ and $\eta$ follows.

Once $\alpha$ is identified, we can further condition on another
event 
\[
G_{2}=\left\{ \tilde{\delta}_{t}\left(\alpha\right)=0\right\} 
\]
where $\tilde{\delta}_{t}\left(\alpha\right)=\left(1+\alpha\right)p_{t-1}-\mu_{t-1}-\alpha\mu_{t}$.
Under the event $G_{2}$, we verify in Appendix Section \ref{sec:Verification}
that  (\ref{eq_Return}) becomes 
\[
R_{t}\left(\theta\right)=\psi\left(\varsigma_{t-1}\right)\tau\Delta_{t-1}=\psi\left(\sqrt{\tau}\frac{1+\alpha}{\alpha\sigma_{x}\sqrt{\eta}}\left|\Delta_{t-1}\right|\right)\tau\Delta_{t-1},
\]
 where $\psi\left(a\right)=2\Lambda\left(a\right)-1$ is strictly
increasing, and non-negative when $a\geq0$. 

Taking the expectation operator $E\left[\left|\cdot\right||G_{2}\right]$
on both sides of the above equation, we have 
\[
E\left[\left|R_{t}\left(\theta\right)\right|\big|G_{2}\right]=\tau E\left[\psi\left(\sqrt{\tau}\frac{1+\alpha}{\alpha\sigma_{x}\sqrt{\eta}}\left|\Delta_{t-1}\right|\right)\left|\Delta_{t-1}\right|\bigg|G_{2}\right].
\]
Since $\left(\alpha,\sigma_{x},\eta\right)$ are already recovered,
in the above equation $\tau$ is the only known parameter. Because
the right-hand side is monotonically increasing in $\tau$ for any
$\tau\geq0$ as long as $\Delta_{t-1}\neq0$, the parameter $\tau$
is identified.

The discussion of identification ensures that we can pin down the
deep parameters from the observable time series given sufficiently
many observations. We proceed to the estimation strategy.

\subsection{Moment Conditions}

Recall that $R_{t}^{\mathrm{r}}$\emph{ }is the\emph{ }real return
and $R_{t}\left(\theta\right)$ is the return according to the model.\emph{
}If the real data is truly generated from the structural model, the
distribution of $\left(R_{t}^{\mathrm{r}}\right)_{t=1}^{T}$ must
be the same as the that of $\left(R_{t}\left(\theta\right)\right)_{t=1}^{T}$.
In reality, the structural model is at best a simplification of the
real world. 

Moment matching is one of the most popular econometric methods to
estimate structural models. We estimate the structural parameter $\theta$
by matching moments of the marginal distribution of returns. First,
as $\sigma_{\mu}$ is identified from the standard deviation of $\varepsilon_{t}^{\mu}$,
we specify the first moment function 
\[
g_{1t}\left(\theta\right)=\left(\varepsilon_{t}^{\mu}\right)^{2}-\sigma_{\mu}^{2},
\]
 since $E\left[T^{-1}\sum_{t=1}^{T}g_{1t}\left(\theta\right)\right]=E\left[T^{-1}\sum_{t=1}^{T}\left(\varepsilon_{t}^{\mu}\right)^{2}\right]-\sigma_{\mu}^{2}=0$.
Next, as the two parameters $\eta$ and $\alpha$ can be identified
given $G_{1}$, we match the conditional mean and variance. Notice
that these two moments are implied by the thin-set identification,
and conditioning on $G_{1}$ literally means selecting only the observations
such that $\Delta_{t-1}=0$. Since $\Delta_{t-1}$ is continuously
distributed, the event $G_{1}$ happens with probability zero. To
avoid the problem of too few local observations, we use a kernel function
to assign weights to each observation, as in \citet{smith2007efficient}
and \citet{gospodinov2012local}. We assign large weights on observations
with small $\left|\Delta_{t-1}\right|$ and small weights on those
with large $\left|\Delta_{t-1}\right|$. Given an appropriate bandwidth
$h_{T}$, we would have enough observations to guarantee the estimation
consistency at $\Delta_{t-1}=0$ asymptotically as $T\to\infty$.
Let $w_{t}^{G_{1}}\left(h_{T}\right)=\phi\left(\Delta_{t-1}/h_{T}\right)$
be the weight of the $t$-th observation, where $h_{T}$ is the bandwidth
and $\phi\left(a\right)=\left(2\pi\right)^{-1/2}\exp\left(-0.5a^{2}\right)$
is the density function of the standard normal. We construct two Gaussian-kernel-weighted
moment functions 
\begin{align*}
g_{2t}\left(\theta\right) & =w_{t}^{G_{1}}\left(h_{T}\right)\left(R_{t}^{\mathrm{r}}-R_{t}\left(\theta\right)\right)\\
g_{3t}\left(\theta\right) & =w_{t}^{G_{1}}\left(h_{T}\right)\left(\left(\tilde{R}_{t}^{\mathrm{r}}\right)^{2}-\tilde{R}_{t}^{2}\left(\theta\right)\right),
\end{align*}
where $\tilde{R}_{t}^{\mathrm{r}}=R_{t}^{\mathrm{r}}-T^{-1}\sum_{t=1}^{T}R_{t}^{\mathrm{r}}$
is the demeaned $R_{t}^{\mathrm{r}}$, and $\tilde{R}_{t}^{2}\left(\theta\right)$
is defined similarly. On the other hand, the chartist parameter $\tau$
is identified conditional on $G_{2}$. The argument for identification
of $\tau$ conditional on $G_{2}$ motivates another kernel-weighted
moment function
\begin{equation}
g_{4t}\left(\theta\right)=w_{t}^{G_{2}}\left(\alpha,h_{T}\right)\left(\left|R_{t}^{\mathrm{r}}\right|-\left|R_{t}\left(\theta\right)\right|\right),\label{eq:g4}
\end{equation}
 where $w_{t}^{G_{2}}\left(\alpha,h_{T}\right)=\phi\left(\tilde{\delta}_{t}\left(\alpha\right)/h_{T}\right)$.
We use the same bandwidth $h_{T}$ in $w_{t}^{G_{1}}\left(h_{T}\right)$
and $w_{t}^{G_{2}}\left(\alpha,h_{T}\right)$ for simplicity. 

Under the assumption that the model is correctly specified, the moments
\[
E\left[\left\{ g_{jt}\left(\theta\right)\right\} _{j=1,\ldots,4}\right]=0_{4\times1}
\]
 pointly identify $\theta$. However, since $\left\{ g_{jt}\left(\theta\right)\right\} _{j=2,3,4}$
are constructed from a sub-population, they only use a small fraction
of the data. As a consequence, the rates of convergence of the kernel-weighted
sample moments are slower than the usual rate of $\sqrt{T}$, so are
the rates of the estimated parameters. It is desirable to improve
the rate of convergence of these parameter estimates by local identification
information.

Following \citet{gagliardini2011efficient} and \citet{antoine2012efficient},
we assume local identification in the sense of \citet{rothenberg1971identification}.
That is, $\theta_{0}$ is \emph{locally identified} if there exists
an open neighborhood of $\theta_{0}$ containing no other $\theta$
that can generate the same distribution. Local identification does
not contradict the thin-set identification. Local identification is
based on the unconditional information of the population. The thin-set
point identification here, however, relies on the conditioning of
two special events that form the sub-population. 

Assuming local identification, we further construct four unconditional
moments with the whole sample. Specifically, we match the mean, variance,
skewness and kurtosis of the returns:
\begin{align*}
g_{5t}\left(\theta\right) & =R_{t}^{\mathrm{r}}-R_{t}\left(\theta\right)\\
g_{6t}\left(\theta\right) & =\left(\tilde{R}_{t}^{\mathrm{r}}\right)^{2}-\tilde{R}_{t}^{2}\left(\theta\right)\\
g_{7t}\left(\theta\right) & =\left(\tilde{R}_{t}^{\mathrm{r}}\right)^{3}-\tilde{R}_{t}^{3}\left(\theta\right)\\
g_{8t}\left(\theta\right) & =\left(\tilde{R}_{t}^{\mathrm{r}}\right)^{4}-\tilde{R}_{t}^{4}\left(\theta\right).
\end{align*}
We focus on these moments, thanks to the well-documented stylized
facts about financial time series, i.e., excessive volatility, negative
skewness, and fat tail in returns \citep{cont2001}. Under local identification,
these unconditional moment functions $\left\{ g_{jt}\left(\theta\right)\right\} _{j=5,\ldots,8}$
improve asymptotic efficiency of the estimator.

The construction of the moments gives a clear interpretation of \emph{indirect
inference} \citep{gourieroux1993indirect}. While $\theta$ is the
deep parameter from the structural model, those eight conditional
and unconditional moments consist of a set of reduced-form parameters.
The principle of indirect inference matches the reduced-form parameters
from the observable data and the counterparts from the structural
model. Model misspecification can be accommodated by indirect inference,
in which the estimated structural parameter $\theta$ is the one that
minimizes some distance between the reduced-form parameter from the
real world and that from the economic theoretical model. Even though
our stylized fundamentalist-chartist model is certainly a simplistic
narrative, the estimation will tune the model to its best approximation
to the features of the observed return time series.

\subsection{Estimation: XMM}

The standard theory of GMM requires that all moments converge at rate
$\sqrt{T}$. Such a premise is violated if we combine the eight moments
$E\left[g_{jt}\left(\theta\right)\right]$, $j=1,\ldots,8$. The unconditional
moments and conditional ones converge to their population means at
different rates. Let $\mathbf{g}_{t}\left(\theta\right)=\left(g_{jt}\left(\theta\right)\right)_{j=1,\ldots,8}$
be the vector of the moment functions. Evaluated at a neighborhood
of the true value, the (scaled) sample unconditional moments $T^{-1/2}\sum_{t=1}^{T}g_{jt}\left(\theta\right)=O_{p}\left(1\right)$
for $j\in\left\{ 1,5,\ldots,8\right\} $, while the (scaled) sample
conditional moments $\left(Th_{T}\right)^{-1/2}\sum_{t=1}^{T}g_{jt}\left(\theta\right)/\sum_{t=1}^{T}w_{t}^{G_{1}}\left(h_{T}\right)=O_{p}\left(1\right)$
for $j\in\left\{ 2,3\right\} $ and $\left(Th_{T}\right)^{-1/2}\sum_{t=1}^{T}g_{4t}\left(\theta\right)/\sum_{t=1}^{T}w_{t}^{G_{2}}\left(\alpha,h_{T}\right)=O_{p}\left(1\right)$.
With such a mixture of sample moments converging at various rates,
the standard asymptotic theory of GMM is inapplicable. Fortunately,
\citet{gagliardini2011efficient} and \citet{antoine2012efficient}
have developed XMM, an extension of GMM, to explicitly incorporate
moments with different rates of convergence. This latest methodological
advancement makes the following empirical estimation possible.

We implement XMM by the continuous updating estimator (CUE) \citep{hansen1996finite}.
 Let $\overline{g}_{j}\left(\theta\right)=T^{-1}\sum_{t=1}^{T}g_{jt}\left(\theta\right)$
be the simple sample average of $\left(g_{jt}\left(\theta\right)\right)_{t=1}^{T}$.
Define the CUE criterion function as
\[
J\left(\theta\right)=T\overline{\mathbf{g}}'\left(\theta\right)\widehat{\Omega}^{-1}\left(\theta\right)\overline{\mathbf{g}}\left(\theta\right),
\]
where $\overline{\mathbf{g}}\left(\theta\right)=\left(\overline{g}_{j}\left(\theta\right)\right)_{j=1}^{8}$,
and $\widehat{\Omega}\left(\theta\right)$ is the sample long-run
variance of $\left(\mathbf{g}_{t}\left(\theta\right)\right)_{t=1}^{T}$.
CUE automates the choice of the weighting matrix so that we do not
have to track the rate of each sample moment, and the scaling factors
in the unconditional moments, $1/\sum_{t=1}^{T}w_{t}^{G_{1}}\left(h_{T}\right)$
and $1/\sum_{t=1}^{T}w_{t}^{G_{2}}\left(\alpha,h_{T}\right)$, are
also canceled out in $\widehat{\Omega}\left(\theta\right)$.

We denote the XMM estimator as\footnote{\label{fn:wG2} While standard kernel weight only depends on $h_{T}$,
here $w_{t}^{G_{2}}\left(\alpha,h_{T}\right)$ also depends on $\alpha$.
In Appendix Section \ref{sec:Verification} we explain that it does
not affect the asymptotic distribution of $\widehat{\theta}_{\mathrm{XMM}}$.}
\begin{equation}
\widehat{\theta}_{\mathrm{XMM}}=\arg\min_{\theta\in\Theta}\ J\left(\theta\right).\label{eq:XMM}
\end{equation}
Under regularity assumptions (see \citet[ p.1203]{gagliardini2011efficient}
or \citet[Theorem 4.3]{antoine2012efficient}), if $h_{T}\to0$, $h_{T}\sqrt{T}\to\infty$
as $T\to\infty$, we have
\begin{equation}
\sqrt{T}\left(\widehat{\theta}_{\mathrm{XMM}}-\theta_{0}\right)\stackrel{\mathrm{d}}{\to}N\left(0,\Sigma\right),\label{eq:asym_normal}
\end{equation}
where $\Sigma$ is the asymptotic variance and it can be consistently
estimated by 
\[
\widehat{\Sigma}=\left[\left(\frac{1}{T}\sum_{t=1}^{T}\frac{\partial}{\partial\theta}\mathbf{g}'_{t}\left(\theta\right)\right)
\widehat{\Omega}^{-1}\left(\theta\right)\left(\frac{1}{T}\sum_{t=1}^{T}\frac{\partial}{\partial\theta'}\mathbf{g}_{t}\left(\theta\right)\right)\right]^{-1}\big|_{\theta=\widehat{\theta}_{\mathrm{XMM}}}.
\]
 Regarding the model specification test, \citet[Theorem 4.4]{antoine2012efficient}
prove that this $J$-statistic still follows the usual $\chi^{2}$
distribution. With eight moments and four unknown parameters, the
degrees of freedom of the $\chi^{2}$ distribution is 4. 

\subsection{Implementation}

We use Robert Shiller's S\&P 500 dataset to construct the price and
the fundamental (Downloadable at \url{http://www.econ.yale.edu/~shiller/data.htm}).
The raw time series $\mathbf{p}^{T}$ is taken as the monthly average
of the daily closing prices, and $\boldsymbol{\mu}^{T}$ is calculated
as the present value of all monthly dividend flows according to the
Gordon growth model \citep{gordon1959dividends}. 

Our discussion of the econometric procedure leaves open several choices
in the implementation. We discuss these issues one by one. The observed
return and the fundamental time series both exhibit upward trends.
We have to filter the trends so that we can focus on the fluctuation
of the stationary time series. Detrending does not change the behavior
of the investors as the growth trend is incorporated in their decision
of the quantity they purchase and the strategy they take. For simplicity,
we fit a linear trend for each time series and then detrend. We find
that the difference in the two trends is very small, which supports
the implication of the efficient market hypothesis that the growth
rate of $\boldsymbol{\mu}^{T}$ and $\mathbf{p}^{T}$ converge in
the long run. We observe that detrending in our data preserves the
pattern of over- and under-pricing periods as the crossing points
of the two raw time series are proximate before and after detrending.
Moreover, for the chartist strategy we need a reference price $p_{t-1}^{\mathrm{ref}}$.
We use a simple 12-month moving average rule $p_{t-1}^{\mathrm{ref}}=\frac{1}{12}\sum_{s=t-12}^{t-1}p_{s}$.

We have assumed that the density function of $\Lambda$ to be symmetric
and its support is the real line. Many distributions satisfy these
conditions, for example the standard normal, the hyperbolic secant
distribution, the Logistic distribution, the Laplace distribution,
and the $t$-distributions of degrees of freedom at least 3 (with
their variance standardized as 1). While $\varepsilon_{it}$ is unobservable,
data provides no guidance about the choice of $\Lambda$. We select
$\Lambda$ as the standard normal for its theoretical and practical
attractiveness. Firstly, under normality the truncated mean function
$\varphi\left(a\right)=-\left(2\pi\right)^{-1/2}\exp\left(-a^{2}/2\right)$
is the (minus) density function of $N\left(0,1\right)$, which is
a built-in function in all modern statistical programming languages.
Secondly, the normal distribution is favorable in justifying the conditional
expectation of $\mu_{t}$ in the fundamental strategy. Given $\mu_{t-1}$
and $x_{it}$, if the fundamentalist takes a prior distribution $\varepsilon_{t}^{\mu}\sim N\left(0,1\right)$,
she will attain the posterior distribution $\mu_{t}|\left(\mu_{t-1},x_{it}\right)\sim N\left(\frac{\mu_{t-1}+\alpha x_{it}}{1+\alpha},\frac{\sigma_{\mu}^{2}}{1+\alpha}\right),$
which delivers \emph{exactly} the weighted average rule for the fundamentalist's
expectation of $\mu_{t}$. 

Throughout this paper, we use the same set of tuning parameters for
all estimation procedures and sample periods. The bandwidth $h_{T}$
in the kernel-weighted sample moments is set as $1.06\widehat{\sigma}_{\Delta}T^{-1/5}$
according to \citet{silverman1986density}'s rule of thumb, where
$\widehat{\sigma}_{\Delta}$ is the sample standard deviation of $\left(\Delta_{t}\right)_{t=1}^{T}$.
The long-run variance is estimated using the Bartlett kernel \citep{newey1987simple};
the number of lags in the kernel is chosen as $1.14\left\lfloor T^{1/3}\right\rfloor $
where $\left\lfloor \cdot\right\rfloor =\max_{b\in\mathbb{N}}\left\{ b\leq\cdot\right\} $,
with the constant and the rate recommended in \citet{andrews1991heteroskedasticity}.
The rates of these tuning parameters satisfy the requirement for the
asymptotic normality, and the estimates are stable in a reasonable
range. 

When applying XMM to the data, we set the compact parameter space
$\Theta$ as $\left[0.001,3\right]^{3}\times\left[0.001,6\right]$,
which is sufficiently wide for $\theta$. We must deal with the local
optimizers in general nonlinear programming. We try many initial values
to enhance the probability of capturing the global minimizer. The
initial value for $\sigma_{\mu}$ is always the sample mean of $\left(\varepsilon_{t}^{\mu}=\mu_{t}-\mu_{t-1}\right)_{t=1}^{T}$.
This sample mean is a consistent estimator, although in theory it
is not as efficient as the XMM estimator since it does not incorporate
the information provided by the other moments. For the other three
parameters $\left(\eta,\tau,\alpha\right)$, the initial value is
independently drawn from the uniform distribution over their parameter
space. Given a randomly generated initial value, we carry out the
nonlinear optimization. We repeat such optimization for 100 times,
save each local minimum, and take the smallest one as the global minimum. 

\section{Empirical Results\label{sec:Empirical-results}}

In this section, we report the empirical results and compare them
with alternative specifications. We first estimate the parameters
with a recent time span from January, 1991 to December, 2013, to which
we refer as \emph{Period 1}. We then repeat the estimation procedure
for two alternative time spans: January, 1961\textemdash December,
1990 (\emph{Period 2}), and January, 1911\textemdash December, 1960
(\emph{Period 3}) for robustness check. 

In Section \ref{subsec:Benchmark:-XMM}, all the eight moments are
incorporated in the estimation, to which we refer as the full model.
Furthermore, we evaluate the effect of the kernel-weighted moments
in Section \ref{subsec:GMM}, and the mixture of the two strategies
in Section \ref{subsec:solo}. 

\subsection{\label{subsec:Benchmark:-XMM} Results from XMM}

\begin{figure}
\begin{centering}
\includegraphics[scale=0.65]{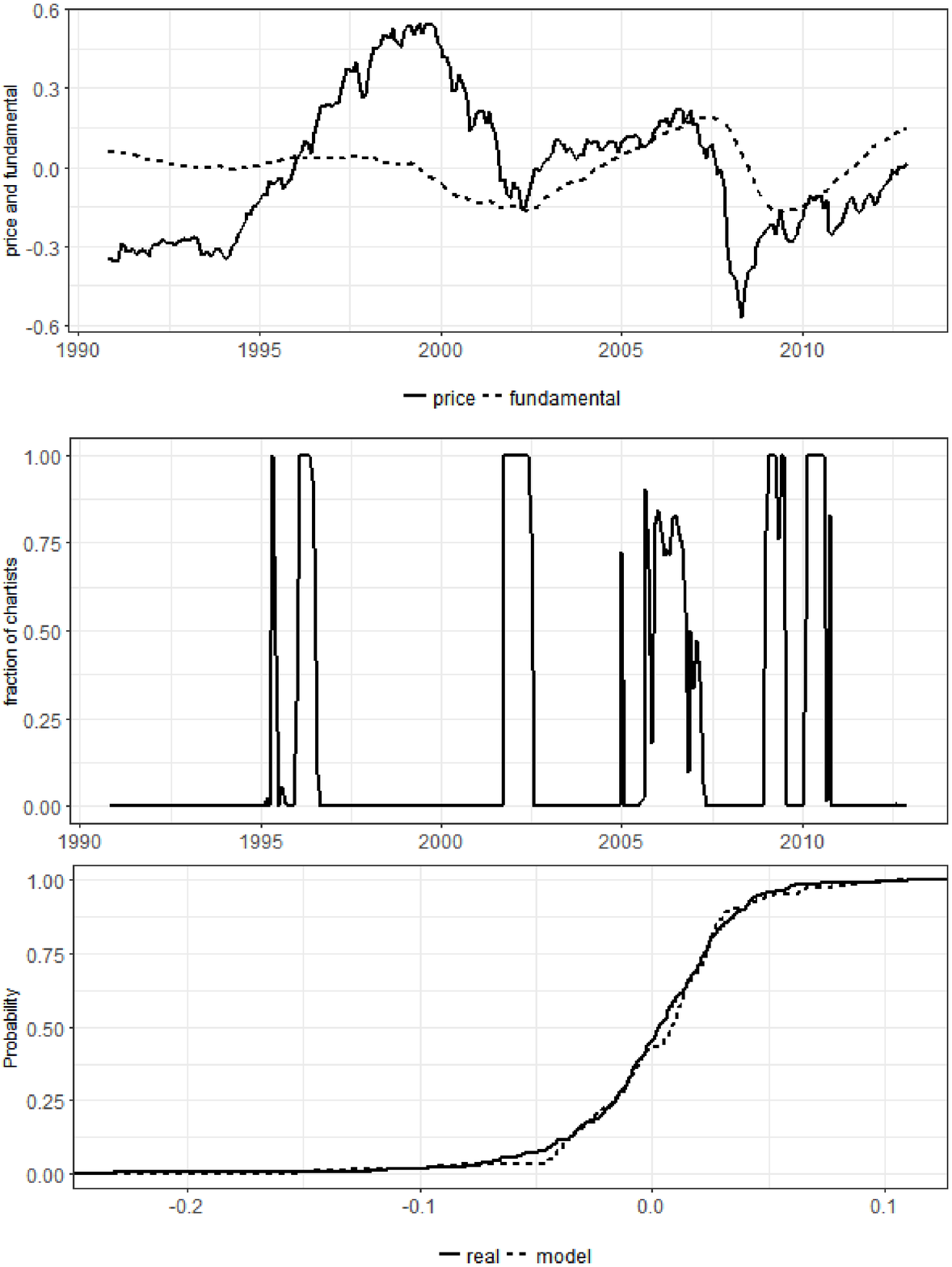}
\par\end{centering}
\caption{\label{fig:plot1} Data, Switching, and Fitting: Period 1 (January,
1991\textendash December, 2013)}

\medskip{}

\small Note: (i) The upper panel shows the linearly detrended price
$\mathbf{p}^{T}$ and fundamental value $\boldsymbol{\mu}^{T}$. (ii)
The middle panel displays the fraction of chartists according to the
full model, computed as $m_{t}(\widehat{\theta}_{\mathrm{XMM}})$.
(iii) The lower panel plots the ECDF of the real time series $R_{t}^{\mathrm{r}}$,
$t=1,\ldots,T$, and that of the predicted $R_{t}(\widehat{\theta}_{\mathrm{XMM}})$,
$t=1,\ldots,T$. Figures \ref{fig:plot2} and \ref{fig:plot3} below
follow the same format.
\end{figure}

The time series of the linearly-detrended price $\mathbf{p}^{T}$
and fundamental value $\boldsymbol{\mu}^{T}$ in Period 1 are shown
in the upper panel of Figure \ref{fig:plot1}. It is apparent that
the price is more volatile than the fundamental. The price sometimes
deviates significantly away from the fundamental value, which corresponds
to boom-bust episodes in the financial history. In the long run, the
price tracks the fundamental value in general, which supports the
market efficiency theory in a long-term perspective.

We take XMM as our benchmark. We report the XMM estimates of $\theta=\left(\sigma_{\mu},\eta,\tau,\alpha\right)$
and the two-sided 95\% asymptotic confidence intervals in Table \ref{tab:tab1-XMM}
for each sample period. All estimates are positive and none of the
confidence intervals contains 0, which is consistent with the economic
interpretation of these parameters. 

The parameters $\eta$ and $\tau$ represent the trading intensity
of the fundamental strategy and the chartist strategy, respectively.
Based on the estimation results of Period 1, the estimate of $\eta$
means that raising the expected return of the fundamental strategy
by 1\% increases the investment flow by 0.10\% on average. On the
other hand, the estimate of $\tau$ implies that 1\% change in the
expected return of the chartist strategy leads to a 0.61\% hike in
the investment flow. The estimate of $\alpha$ is $1.71$ suggests
that investors update their expected fundamental value aggressively
by overweighing the private information relative to the common prior
on the historical fundamental value, as the private information is
more precise than the public information. In terms of the model specification
test, the $J$-statistic is 2.22 with the $p$-value 0.70. It does
not reject the model, indicating that our model can be a reasonable
description of the data generating process for this sample period.

\begin{table}[h]
\caption{\label{tab:tab1-XMM}Estimation Results of XMM for the Full Model}

\bigskip{}
\begin{centering}
\begin{tabular}{c|cc||c|cc||c|cc||c}
\hline 
 & \multicolumn{3}{c|}{Period 1} & \multicolumn{3}{c|}{Period 2} & \multicolumn{3}{c}{Period 3}\tabularnewline
 & est. & \multicolumn{2}{c|}{95\% CI } & est. & \multicolumn{2}{c|}{95\% CI } & est. & \multicolumn{2}{c}{95\% CI }\tabularnewline
\hline 
$\sigma_{\mu}$ & 0.014 & \multicolumn{2}{c|}{(0.008, 0.020)} & 0.007 & \multicolumn{2}{c|}{(0.006, 0.008)} & 0.030 & \multicolumn{2}{c}{(0.024, 0.037)}\tabularnewline
$\eta$ & 0.102 & \multicolumn{2}{c|}{(0.073, 0.131)} & 0.171 & \multicolumn{2}{c|}{(0.141, 0.202)} & 0.224 & \multicolumn{2}{c}{(0.170, 0.277)}\tabularnewline
$\tau$ & 0.612 & \multicolumn{2}{c|}{(0.390, 0.835)} & 0.758 & \multicolumn{2}{c|}{(0.607, 0.908)} & 1.099 & \multicolumn{2}{c}{(0.794, 1.405)}\tabularnewline
$\alpha$ & 1.713 & \multicolumn{2}{c|}{(0.688, 2.739)} & 2.021 & \multicolumn{2}{c|}{(1.127, 2.915)} & 3.894 & \multicolumn{2}{c}{(2.401, 5.387)}\tabularnewline
\emph{J}-stat. & \multicolumn{3}{c|}{$2.222$ } & \multicolumn{3}{c|}{7.202 } & \multicolumn{3}{c}{$7.471$}\tabularnewline
\emph{p}-value & \multicolumn{3}{c|}{(0.695)} & \multicolumn{3}{c|}{(0.126)} & \multicolumn{3}{c}{ (0.113)}\tabularnewline
\hline 
\end{tabular}
\par\end{centering}
\bigskip{}
\raggedright{}\small Note: Each column represents the sample period
for estimation, where Period 1 is January, 1991\textemdash December,
2013, Period 2 is January, 1961\textemdash December, 1990, and Period
3 is January, 1911\textemdash December, 1960. For each sample period,
this table displays the point estimates (est.) of the four structural
parameters and the 95\% asymptotic confidence intervals (95\% CI).
The $J$-statistic ($J$-stat.) of the over-identification test and
the corresponding $p$-value are also reported. Under the null hypothesis
of correct moment specification, the $J$-statistics asymptotically
follows $\chi^{2}\left(4\right)$. Tables \ref{tab:tab1-GMM} and
\ref{tab:solo strategy} below follow the same format. 
\end{table}

Besides the values of the structural parameters, we are also interested
in the endogenous switching of the financial agents between the chartist
and fundamental strategies. It is illustrated in the middle panel
of Figure \ref{fig:plot1}. Consistent with the model's prediction,
 the market is dominated by fundamentalists when the asset is excessively
mispriced, and by chartists when the price moves more closely around
the fundamental value. How agents switched between the heterogeneous
strategies during the recent global financial crisis is of particular
interest. When the market was booming during 2005\textendash 2007,
many agents clustered to be chartists, who traded on price trends.
When the trend was reversed in late 2007, the market fraction of chartists
declined sharply. Fundamentalists prevailed the market in 2008\textendash 2009,
the most volatile years during the global financial crisis. In that
episode, financial assets were overwhelmingly underpriced (as illustrated
in the upper panel of Figure \ref{fig:plot1}), and fundamentalists
had accumulated strong buying force that drove the price up toward
its fundamental value. However, the price may not converge to the
fundamental value immediately after the fundamentalists occupy the
market. The presence of information friction produces such inertia
in our model. No similar patterns of switching was found during the
dot-com crisis. In the early period of dot-com bubble formation, chartists
dominated the market. As the asset became more and more overpriced,
agents switched to fundamentalists. The bubble continued to grow even
after fundamentalists fully occupied the market. In a highly noisy
environment, some fundamentalists might wrongly extrapolate the asset
to be underpriced even if it was actually overpriced. 

\begin{table}[h]
\caption{\label{tab:predicted_moments}Sample Moments of Real Return and Fitted
Return}

\bigskip{}
\begin{centering}
\begin{tabular}{llrrrr}
\hline 
 &  & Mean & Stan. Dev. & Skewness & Kurtosis\tabularnewline
\hline 
 & Real return & 0.000 & 0.037 & -1.380 & 9.280\tabularnewline
 & XMM full model & 0.002 & 0.036 & -0.918 & 6.550\tabularnewline
Period 1 & GMM  & 0.001 & 0.036 & -1.180 & 9.020\tabularnewline
 & XMM fundamentalist-only & 0.000 & 0.029 & -0.291 & 2.460\tabularnewline
 & XMM chartist-only & 0.000 & 0.089 & -1.450 & 5.990\tabularnewline
\hline 
 & Real return & 0.000 & 0.036 & -0.786 & 4.970\tabularnewline
 & XMM full model & 0.000 & 0.041 & -0.214 & 4.340\tabularnewline
Period 2 & GMM  & -0.001 & 0.037 & -0.435 & 5.090\tabularnewline
 & XMM fundamentalist-only & 0.000 & 0.028 & 0.261 & 2.330\tabularnewline
 & XMM chartist-only & 0.000 & 0.107 & -0.670 & 3.310\tabularnewline
\hline 
 & Real return & 0.000 & 0.051 & -0.175 & 15.300\tabularnewline
 & XMM full model & 0.002 & 0.060 & -0.494 & 11.500\tabularnewline
Period 3 & GMM  & 0.001 & 0.050 & -0.177 & 13.200\tabularnewline
 & XMM fundamentalist-only & 0.000 & 0.038 & 0.347 & 2.810\tabularnewline
 & XMM chartist-only & 0.000 & 0.141 & -1.130 & 6.680\tabularnewline
\hline 
\end{tabular}
\par\end{centering}
\bigskip{}
\raggedright{}\small  Note: This table displays the mean, standard
deviation (Stan. Dev.), and the standardized skewness and kurtosis
of the real returns and the predicted returns. For a sample $x_{1},\ldots,x_{T}$,
the standardized skewness and kurtosis here are respectively computed
as $\widehat{\sigma}^{-3}T^{-1}\sum_{t=1}^{T}\left(x_{t}-\bar{x}\right)^{3}$
and $\widehat{\sigma}^{-4}T^{-1}\sum_{t=1}^{T}\left(x_{t}-\bar{x}\right)^{4}$,
where $\bar{x}$ and $\widehat{\sigma}$ are the sample mean and standard
deviation. In each period, the moments of the real returns are computed
from the observed time series $R_{t}^{\mathrm{r}}$, $t=1,\ldots,T$,
while the other rows are calculated from $R_{t}(\widehat{\theta})$,
$t=1,\ldots,T$, where $\widehat{\theta}$ is the corresponding estimate. 
\end{table}

To examine the performance of moment matching, we plug in the estimated
parameters into the model to predict the return. In the lower panel
of Figure \ref{fig:plot1}, the solid line is the empirical cumulative
distribution function (ECDF) of the real data $\left(R_{t}^{\mathrm{r}}\right)_{t=1}^{T}$,
and the dashed line is the ECDF of the fitted return series $\left(R_{t}\left(\widehat{\theta}_{\mathrm{XMM}}\right)\right)_{t=1}^{T}$.
The two ECDF curves closely track each other. As shown in the upper
panel of Table \ref{tab:predicted_moments}, the predicted returns
generated from XMM have a mean return close to zero, a variance around
0.04, a negative skewness, and a kurtosis that is larger than 3. These
sample moments are similar to those of the real return series.

\begin{figure}
\begin{centering}
\includegraphics[scale=0.65]{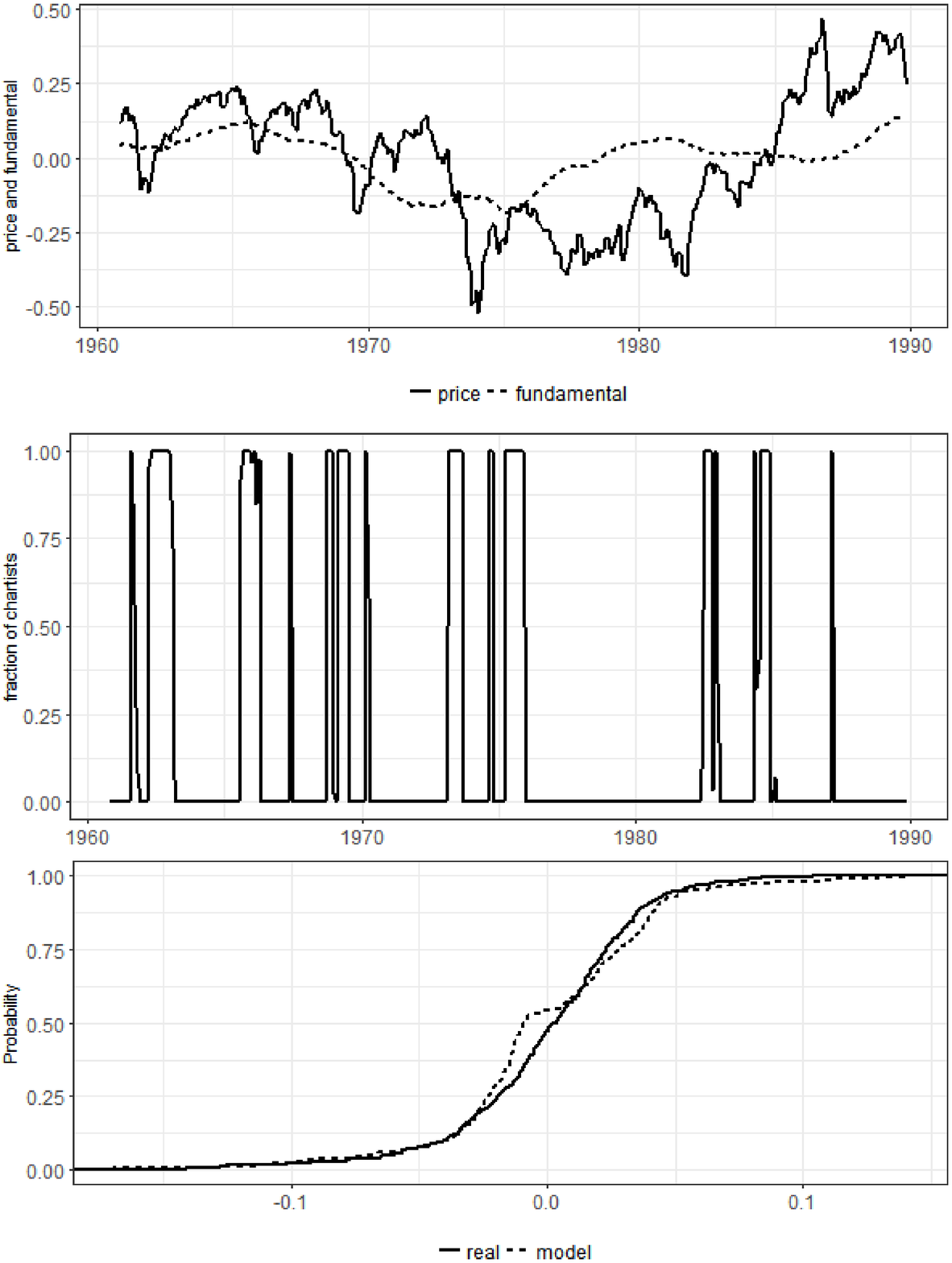}
\par\end{centering}
\caption{\label{fig:plot2} Data, Switching, and Fitting: Period 2 (January,
1961\textendash December, 1990)}
\end{figure}

Next, we repeat the same exercises for other sample periods to check
the robustness of the empirical results. Figure \ref{fig:plot2} and
the second column in Table \ref{tab:tab1-XMM} display the results
of Period 2. Again, the $J$-statistic does not reject the model specification.
The estimated coefficients are comparable with those in Period 1.
The fitted returns match well with the real data in terms of ECDF
and the four moments, as shown in the middle panel of Table \ref{tab:predicted_moments}.
Moreover, consistent with the previous results, we observe from the
middle panel of Figure \ref{fig:plot2} that chartists prevailed when
the asset was moderately priced, for example in 1976, while fundamentalists
dominated the market when the price deviated significantly away from
the fundamental, for example in 1978\textendash 1982.

\begin{figure}
\begin{centering}
\includegraphics[scale=0.65]{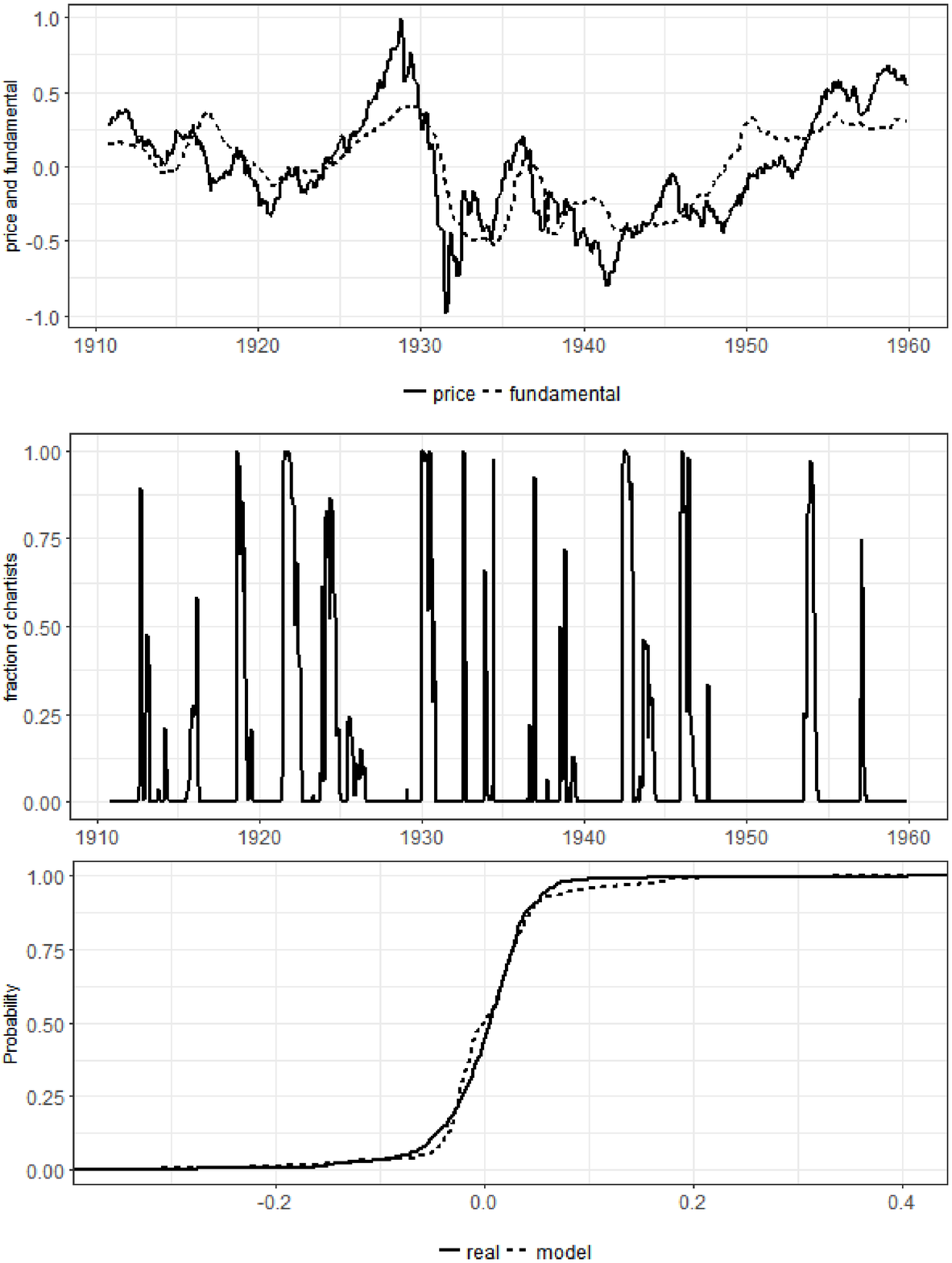}
\par\end{centering}
\caption{\label{fig:plot3} Data, Switching, and Fitting: Period 3 (January,
1910\textendash December, 1960)}
\end{figure}

Figure \ref{fig:plot3} and the third column of Table \ref{tab:tab1-XMM}
report the results for Period 3, a half century that witnessed the
Great Depression. The high volatility in this era is manifest as shown
in Table \ref{tab:predicted_moments}, with a kurtosis of 15.30 in
the real return, the largest among the three sample periods. In terms
of the point estimates, the scale of the estimated coefficients $\tau$
and $\eta$ are larger than those reported in the other two periods,
showing that both fundamentalists and chartists responded more sensitively
to the expected returns. The estimated coefficient $\alpha$ in Period
3 is about twice as large as that in Period 1 or 2, suggesting fundamentalists
updated information more aggressively in response to the volatile
market. In Figure \ref{fig:plot3} we again observe the switching
from chartists to fundamentalists when the asset was excessively mispriced
and from fundamentalists to chartists when the asset was moderately
mispriced. 

In the following sections, we estimate some simple alternative models
and compare the empirical results with those discussed in this section.

\subsection{\label{subsec:GMM} Estimation with Unconditional Moments Only: GMM}

The standard GMM utilizes only the unconditional moments in estimation.
For comparison, we implement GMM (CUE) with the  moment functions
$\left\{ g_{jt}\left(\theta\right)\right\} _{j=1,5,6,7,8}$, and the
results are reported in Table \ref{tab:tab1-GMM}. Ignoring $\left\{ g_{jt}\left(\theta\right)\right\} _{j=2,3,4}$,
which contains information from the theoretical model, weakens the
asymptotic efficiency of parameter estimation. In our context, such
efficiency loss is reflected in the confidence intervals\textemdash in
most cases the confidence intervals of the GMM estimator are wider
than their XMM counterparts. In particular, the confidence interval
of $\eta$ includes 0, which is highly undesirable since the identification
of the parameters relies on a positive $\eta$. In contrast, when
conditional moments are accounted for, the confidence intervals of
$\eta$ are clearly deviated away from 0 (see Table \ref{tab:tab1-XMM}).
In the meantime, with fewer restrictions GMM improves the in-sample
fitting. The model emulates the data more closely in terms of moment
matching, as shown in Table \ref{tab:predicted_moments} with the
kurtosis of the predicted return closer to that of the real data. 

\begin{table}[h]
\caption{\label{tab:tab1-GMM}Estimation Results of GMM with the Unconditional
Moments}

\bigskip{}
\begin{centering}
\begin{tabular}{c|cc||c|cc||c|cc||c}
\hline 
 & \multicolumn{3}{c|}{Period 1} & \multicolumn{3}{c|}{Period 2} & \multicolumn{3}{c}{Period 3}\tabularnewline
 & est. & \multicolumn{2}{c|}{95\% CI } & est. & \multicolumn{2}{c|}{95\% CI } & est. & \multicolumn{2}{c}{95\% CI }\tabularnewline
\hline 
$\sigma_{\mu}$ & 0.014 & \multicolumn{2}{c|}{(0.010, 0.019)} & 0.002 & \multicolumn{2}{c|}{(-0.333, 0.337)} & 0.029 & \multicolumn{2}{c}{(0.023, 0.036)}\tabularnewline
$\eta$ & 0.109 & \multicolumn{2}{c|}{(-0.020, 0.238)} & 0.055 & \multicolumn{2}{c|}{(-0.220, 0.330)} & 0.280 & \multicolumn{2}{c}{(-0.061, 0.621)}\tabularnewline
$\tau$ & 0.676 & \multicolumn{2}{c|}{(0.359, 0.992)} & 0.702 & \multicolumn{2}{c|}{(0.502, 0.902)} & 0.685 & \multicolumn{2}{c}{(-0.028, 1.398)}\tabularnewline
$\alpha$ & 2.513 & \multicolumn{2}{c|}{(0.613, 4.413)} & 3.363 & \multicolumn{2}{c|}{(1.405, 5.321)} & 2.765 & \multicolumn{2}{c}{(0.823, 4.707)}\tabularnewline
\emph{J}-stat. & \multicolumn{3}{c|}{0.035} & \multicolumn{3}{c|}{0.082} & \multicolumn{3}{c}{0.066}\tabularnewline
\emph{p}-value & \multicolumn{3}{c|}{(0.852)} & \multicolumn{3}{c|}{(0.774)} & \multicolumn{3}{c}{ (0.797)}\tabularnewline
\hline 
\end{tabular}
\par\end{centering}
\bigskip{}
\raggedright{}\small Note: Similar to Table \ref{tab:tab1-XMM},
this table displays the point estimates (est.) and the 95\% asymptotic
confidence intervals (95\% CI) for each sample period. Since only
the unconditional moments are used in the estimation, the $J$-statistics
of the over-identification test follows $\chi^{2}\left(1\right)$
asymptotic distribution under the null.
\end{table}

\subsection{\label{subsec:solo} Estimation with a Solo Strategy}

A common feature of the strategy switching in Figure \ref{fig:plot1}\textendash \ref{fig:plot3}
is that fundamentalists dominate the market more frequently than chartists.
This observation raises the question of the necessity of introducing
the two investment strategies to characterize the price movement.
This section explores whether a solo-strategy model is sufficient
to capture the price dynamics. 

The fundamentalist-only model is a sub-model of the two-strategy benchmark
model. When $\tau=0$ and $\eta>0$, the chartist strategy generates
zero profit so that no investor will adopt it. The predicted return
of the fundamentalist-only model follows  (\ref{eq:fun_only}). Since
the kernel-weighted moment functions $g_{jt}\left(\theta\right),$
$j\in\left\{ 2,3,4\right\} $, remain valid in the sub-model, we estimate
the parameters $\left(\sigma_{\mu},\eta,\alpha\right)$ by XMM with
the same eight moments as in  (\ref{eq:XMM}) but setting $\tau=0$.
With the restriction $\tau=0$, the $J$-statistic follows $\chi^{2}\left(5\right)$
asymptotically under the null.

We report the empirical results in the upper panel of Table \ref{tab:solo strategy}.
The estimates stay positive and statistically significant as the 95\%
confidence intervals are all above 0. This is consistent with the
results from the full model and provides evidence of the presence
of the fundamentalist trading in the market. Under the null hypothesis
that the fundamentalist-only model is correctly specified, the $J$-statistics
are 15.74, 16.44 and 32.15 in Period 1\textendash 3, respectively,
which are associated with $p$-value less than 1\%. The strong rejection
means that the fundamental strategy solely is incapable of mimicking
the observed price movements. Moreover, in Table \ref{tab:predicted_moments}
the moments of fitted returns are far away from the real ones. In
particular, the fitted kurtosis is less than 3 throughout the three
sample periods, which contradicts the fat-tail phenomenon observed
in the real data.

\begin{table}[h]
\caption{\label{tab:solo strategy} Estimation Results of the Solo Strategy
Models}

\bigskip{}
\begin{centering}
\par\end{centering}
\begin{centering}
\begin{tabular}{c|cc||c|cc||c|cc||c}
\hline 
 & \multicolumn{3}{c}{Period 1} & \multicolumn{3}{c}{Period 2} & \multicolumn{3}{c}{Period 3}\tabularnewline
 & est. & \multicolumn{2}{c}{95\% CI} & est. & \multicolumn{2}{c}{95\% CI} & est. & \multicolumn{2}{c}{95\% CI}\tabularnewline
\hline 
 & \multicolumn{9}{c}{Fundamentalist-only}\tabularnewline
$\sigma_{\mu}$ & 0.015 & \multicolumn{2}{c|}{(0.010, 0.020)} & 0.007 & \multicolumn{2}{c|}{(0.006, 0.008)} & 0.030 & \multicolumn{2}{c}{(0.024, 0.036)}\tabularnewline
$\eta$ & 0.084 & \multicolumn{2}{c|}{(0.060, 0.108)} & 0.157 & \multicolumn{2}{c|}{(0.125, 0.189)} & 0.182 & \multicolumn{2}{c}{(0.130, 0.233)}\tabularnewline
$\alpha$ & 0.612 & \multicolumn{2}{c|}{(0.378, 0.846)} & 1.178 & \multicolumn{2}{c|}{(0.826, 1.530)} & 1.239 & \multicolumn{2}{c}{(0.797, 1.682)}\tabularnewline
\emph{J}-stat. & \multicolumn{3}{c|}{$15.740$} & \multicolumn{3}{c|}{$16.442$} & \multicolumn{3}{c}{$32.151$}\tabularnewline
\emph{p}-value & \multicolumn{3}{c|}{ (0.008)} & \multicolumn{3}{c|}{ (0.006)} & \multicolumn{3}{c}{ (0.000)}\tabularnewline
\hline 
 & \multicolumn{9}{c}{Chartist-only}\tabularnewline
$\sigma_{\mu}$ & 0.009 & \multicolumn{2}{c|}{(0.002, 0.016)} & 0.007 & \multicolumn{2}{c|}{(0.006, 0.008)} & 0.023 & \multicolumn{2}{c}{(0.016, 0.031)}\tabularnewline
$\tau$ & 0.871 & \multicolumn{2}{c|}{(0.792, 0.949)} & 1.179 & \multicolumn{2}{c|}{(1.071, 1.286)} & 1.052 & \multicolumn{2}{c}{(0.944, 1.161)}\tabularnewline
\emph{J}-stat. & \multicolumn{3}{c|}{$37.697$} & \multicolumn{3}{c|}{$94.914$} & \multicolumn{3}{c}{$44.639$}\tabularnewline
\emph{p}-value & \multicolumn{3}{c|}{(0.000)} & \multicolumn{3}{c|}{(0.000)} & \multicolumn{3}{c}{(0.000)}\tabularnewline
\hline 
\end{tabular}\bigskip{}
\par\end{centering}
\small Note: Under the null hypothesis, the $J$-statistic of the
fundamentalist-only model follows $\chi^{2}\left(5\right)$ asymptotic
distribution, and that of the chartist-only model is $\chi^{2}\left(4\right)$
asymptotically. The corresponding $p$-values are so small that the
over-identification tests are rejected in all cases at 1\% size.
\end{table}

If the fundamentalist-only model is insufficient to capture the real
return, how about the chartist-only model? The lower panel of Table
\ref{tab:solo strategy} displays the XMM estimation results across
the three sample periods.\footnote{A formal test of the chartist-only model is more complicated than
the fundamentalist-only model because the full model precludes $\eta=0$
due to its presence in the denominator in $\zeta_{t-1}$. The implementation
is detailed in Supplement Section S2.} The $J$-statistics clearly reject the chartist-only model at any
commonly used test size, and the moment matching in Table \ref{tab:predicted_moments}
is poorer than the full model.

In view of the empirical results, neither the fundamentalist strategy
nor the chartist strategy alone reasonably matches the data. The mixture
of the two trading strategies is effective in improving the model
fitting. 

\section{Conclusion\label{sec:Conclusion}}

In this paper, we develop a structural asset pricing model with information-driven
behavioral heterogeneity. For this highly nonlinear model, we formally
identify the structural parameters via thin-set identification. The
thin-set identification and the follow-up estimation techniques are
applicable to other heterogeneous agent models involving a mixture
of investment strategies. 

We estimate the parameters by XMM, and conduct inference for the model
specification. The empirical results show that the structural model
emulates the S\&P 500 index. Investors switch between the fundamental
and chartist strategies evolutionarily in response to the dynamic
market conditions. Agents tend to cluster toward the chartist strategy
when the market environment waxes and wanes, and their collective
trading actions cause substantial asset mispricing that sometimes
turn into bubbles and crashes. However, when the asset is significantly
overpriced or underpriced, agents tend to revert to the fundamental
strategy, which corrects the mispricing and restores the market efficiency.
The switching is found to be crucial for the empirical fitness of
the structural model. Models with only one strategy significantly
underperform the structural model in terms of matching the real price
movement. 

In this Appendix of this paper, we present the step-by-step development
of the structural model as well as the derivation of some technical
claims in the main text. Moreover, we provide an Online Supplement
for additional empirical results, extension, and implementation. \newpage{}
\begin{center}
{\huge{}Appendix}
\par\end{center}{\huge \par}

\begin{appendices}

\section{\label{sec:Complete-Description-of } Complete Description of the
Structural Model}

This section describes the information-based structural model, summarized
in Section \ref{sec:Information-Based-Structural-Mod}, step by step.

\subsection{Investment Strategies\label{subsec:Investment-Strategies}}

In each period, the $f$-advisor updates the expected mean of $\mu_{t}$
after learning the private information $x_{it}$ from the agent $i$.
She follows a weighted average rule 
\[
E_{it-1}^{f}\left[\mu_{t}\right]=\frac{\mu_{t-1}/\sigma_{\mu}^{2}+x_{it}/\sigma_{x}^{2}}{1/\sigma_{\mu}^{2}+1/\sigma_{x}^{2}}=\frac{\mu_{t-1}+\alpha x_{it}}{1+\alpha},
\]
where the weight is the information precision (the inverse of variance),
$\alpha=\sigma_{\mu}^{2}/\sigma_{x}^{2}$ is the precision of private
information relative to public information, and $E_{it-1}^{f}\left[\cdot\right]=E\left[\cdot|x_{it},\mathbf{p}^{t-1},\boldsymbol{\mu}^{t-1}\right]$
is the expectation of the $f$-advisor conditional on the past public
information as well as the private signal $x_{it}$. She believes
in the efficient market hypothesis under which the price tracks the
fundamental value. Let her \emph{perceived} return be $R_{t}^{f}=p_{t}^{f}-p_{t-1}$,
where $p_{t}^{f}$ is the \emph{perceived} price to be realized in
time $t$. She expects the period-$t$ return to be
\[
E_{it-1}^{f}\left[R_{t}^{f}\right]=E_{it-1}^{f}\left[p_{t}^{f}\right]-p_{t-1}=E_{it-1}^{f}\left[\mu_{t}\right]-p_{t-1}=\frac{\mu_{t-1}+\alpha x_{it}}{1+\alpha}-p_{t-1}=\frac{\alpha\sigma_{x}}{1+\alpha}\left(\varepsilon_{it}-\delta_{t}\right),
\]
where the second equality is implies by the efficient market hypothesis,
and the last equality by the definitions of $x_{it}$ and $\delta_{t}$. 

The $c$-advisor, on the other hand, employs technical analysis to
forecast the price movements. She ignores the private information
$x_{it}$ and the fundamental $\boldsymbol{\mu}^{t-1}$, even though
they are accessible. Let $E_{t-1}^{c}\left[\cdot\right]=E^{c}\left[\cdot|\mathbf{p}^{t-1}\right]$
be the $c$-advisor's expectation conditional on past prices. The
$c$-advisor believes that the past price trend captured by $\Delta_{t-1}$
would persist in the following period. Let $p_{t}^{c}$ be the chartist's
perceived price to be realized at time $t$, and $R_{t}^{c}=p_{t}^{c}-p_{t-1}$
be the perceived return. Her expected period-$t$ return is
\[
E_{t-1}^{c}\left[R_{t}^{c}\right]=E_{t-1}^{c}\left[p_{t}^{c}|\mathbf{p}^{t-1}\right]-p_{t-1}=\Delta_{t-1}.
\]

{} Let $f$-advisor's expected utility $E^{f}\left[U\right]=-\exp\left(-A^{f}\left(\mu_{W}-\frac{A^{f}}{2}\sigma_{W}^{2}\right)\right)$,
where $\mu_{W}=E\left[W\right]$ and $\sigma_{W}^{2}=\mathrm{var}\left[W\right]$
is the mean and variance of the wealth $W$, respectively, and $A^{f}>0$
is a constant.\footnote{Such a functional form can be formally derived under the constant
absolute risk aversion (CARA) exponential utility function $U=-\exp\left(-A^{f}\cdot W\right)\text{,}$
where $A^{f}>0$ is the absolute risk aversion coefficient, and $W\sim N\left(\mu,\sigma^{2}\right)$
is normally distributed. Such a CARA utility function takes into account
the trade-off between risk and return and it facilitates \textcolor{black}{mean-variance
analysis. It is widely used in the literature, for example \citet{Barberis2016}.}} Maximizing this utility is essentially maximizing $\mu_{W}-\frac{A^{f}}{2}\sigma_{W}^{2}$,
the difference between the mean and the variance multiplied by a constant.

Given the dynamics of wealth growth $W_{it}=W_{it-1}+q_{it}^{f}R_{t}$,
we apply the expected utility function to the $f$-advisor at the
beginning of time $t$:
\begin{align*}
E_{it-1}^{f}\left[U_{it}\right] & =-\exp\left(-A^{f}\left(W_{it-1}+q_{it}^{f}E_{it-1}^{f}[R_{t}^{f}]-\frac{A^{f}}{2}\left(q_{it}^{f}\right)^{2}\mbox{var}{}_{it-1}^{f}[R_{t}^{f}]\right)\right).
\end{align*}
The $f$-advisor who maximizes the expected utility recommends the
optimal investment flow 
\begin{equation}
q_{it}^{f*}=E_{it-1}^{f}[R_{t}^{f}]/\left(A^{f}\cdot\mbox{var}{}_{it-1}^{f}[R_{t}^{f}]\right)=\eta\frac{\alpha\sigma_{x}}{1+\alpha}\left(\varepsilon_{it}-\delta_{t}\right),\label{eq:q_f}
\end{equation}
where $\eta=1/\left(A^{f}\cdot\mbox{var}{}_{it-1}^{f}[R_{t}^{f}]\right)$.
We assume that $\mbox{var}{}_{it-1}^{f}[R_{t}^{f}]$ is a constant
independent of $i$ and $t$. Similar expected utility analysis applies
to the $c$-advisor, whose expected utility $E^{c}\left[U\right]=-\exp\left(-A^{c}\left(\mu_{W}-\frac{A^{c}}{2}\sigma_{W}^{2}\right)\right)$.
As a result, the $c$-advisor recommends the optimal investment flow
\begin{equation}
q_{t}^{c*}=E_{t-1}^{c}\left[R_{t}^{c}\right]/\left(A^{c}\cdot\mbox{var}{}_{t-1}^{c}\left[R_{t}^{c}\right]\right)=\tau\Delta_{t-1},\label{eq:q_c}
\end{equation}
where $\tau=1/\left(A^{c}\cdot\mbox{var}{}_{t-1}^{c}\left[R_{t}^{c}\right]\right)$
as $\mbox{var}{}_{t-1}^{c}\left[R_{t}^{c}\right]$ is assumed to be
a constant. 

Both the conditional variances of the return are assumed time-invariant
for the following reasons.\footnote{In Supplement Section S4, we discuss the possibility of extending
the theoretical model to allow individual- and/or time-varying conditional
variances and its implications to identification and estimation.} (i) The traders follow naive investment rules so that their perceived
$R_{t}^{f}$ and $R_{t}^{c}$ are not directly observable and the
conditional variance cannot be estimated from the data. In such a
setup, the constant conditional variance is a convenient assumption
following the literature, for example \citet[p.1239]{brock1998heterogeneous}
and \citet[p.32]{Barberis2016}, among many others cited in this paper.
(ii) The constant conditional variance allows us to derive a simple
explicit form for the price dynamics, which simplifies the estimation.

\subsection{Choice of Strategies and Aggregation of Demand \label{subsec:Choice-of-Strategies}}

Financial advisors advocate optimal investment flows $q_{it}^{f*}$
and $q_{t}^{c*}$ based on their independent analysis. Each agent
is well informed of both strategies as well as the rationale behind
(\ref{eq:q_f}) and (\ref{eq:q_c}). The agent takes only one of the
two strategies. The expected profit from a strategy is the product
of the expected return and investment flow, i.e., 
\[
\pi_{it}^{f}=q_{it}^{f*}\frac{\alpha\sigma_{x}}{1+\alpha}\left(\varepsilon_{it}-\delta_{t}\right)=\eta\left(\frac{\alpha\sigma_{x}}{1+\alpha}\left(\varepsilon_{it}-\delta_{t}\right)\right)^{2}
\]
 for the fundamental strategy, and 
\[
\pi_{t}^{c}=q_{t}^{c*}\Delta_{t-1}=\tau\Delta_{t-1}^{2}
\]
 for the chartist strategy. The agent prioritizes investment profitability
and chooses the strategy that yields a higher expected profit. Let
$\bar{\varepsilon}_{t}$ be the threshold such that $\pi_{it}^{f}=\pi_{t}^{c}$
when $\varepsilon_{it}=\bar{\varepsilon}_{t}$. We solve $\eta\left(\frac{\alpha\sigma_{x}}{1+\alpha}\left(\bar{\varepsilon}_{t}-\delta_{t}\right)\right)^{2}=\tau\Delta_{t-1}^{2}$
to obtain 
\[
\bar{\varepsilon}_{t}=\delta_{t}\pm\frac{1+\alpha}{\alpha\sigma_{x}}\sqrt{\frac{\tau}{\eta}}\left|\Delta_{t-1}\right|=\delta_{t}\pm\zeta_{t-1},
\]
and the lower bound $\bar{\varepsilon}_{t}^{m}=\delta_{t}-\zeta_{t-1}$
and upper bound $\bar{\varepsilon}_{t}^{M}=\delta_{t}+\zeta_{t-1}$
follow.

Since $\pi_{it}^{f}$ is a convex function of $\varepsilon_{it}$
while $\pi_{t}^{c}$ is independent of $\varepsilon_{it}$, we have
$\pi_{it}^{f}<\pi_{t}^{c}$ if $\varepsilon_{it}\in(\bar{\varepsilon}_{t}^{m},\bar{\varepsilon}_{t}^{M})$.
The agent, who seeks to maximize her expected profit, acts on the
chartist strategy if $\varepsilon_{it}\in(\bar{\varepsilon}_{t}^{m},\bar{\varepsilon}_{t}^{M})$,
whereas she carries out the fundamental strategy otherwise. When $\pi_{it}^{f}=\pi_{t}^{c}$,
the agent would be indifferent between the two strategies, in which
case we assume she adopts the fundamental strategy. As a result, the
individual investment flow is 
\begin{equation}
q_{it}^{*}=q_{it}^{f*}\cdot\mathbf{1}\left\{ \varepsilon_{it}\in(-\infty,\bar{\varepsilon}_{t}^{m}]\cup[\bar{\varepsilon}_{t}^{M},\infty)\right\} +q_{t}^{c*}\cdot\mathbf{1}\left\{ \varepsilon_{it}\in(\bar{\varepsilon}_{t}^{m},\bar{\varepsilon}_{t}^{M})\right\} ,\label{eq:q_it}
\end{equation}
where $\boldsymbol{1}\left\{ \cdot\right\} $ is the indicator function. 

In the market, the fraction of chartists is given by $m_{t}=\Lambda\left(\bar{\varepsilon}_{t}^{M}\right)-\Lambda\left(\bar{\varepsilon}_{t}^{m}\right)$,
and the fraction of fundamentalists is $1-m_{t}$. Conditional on
the past information and $\mu_{t}$, the aggregate demand of all agents
is 
\begin{eqnarray*}
D_{t}\left(\theta\right) & = & \int_{-\infty}^{\infty}q_{it}^{*}d\Lambda\left(\varepsilon_{it}\right)\\
 & = & \int_{(-\infty,\bar{\varepsilon}_{t}^{m}]\cup[\bar{\varepsilon}_{t}^{M},\infty)}\frac{\eta\alpha\sigma_{x}}{1+\alpha}\left(\varepsilon_{it}-\delta_{t}\right)d\Lambda\left(\varepsilon_{it}\right)+\tau m_{t}\Delta_{t-1}\\
 & = & \frac{\eta\alpha\sigma_{x}}{1+\alpha}\left(\int_{-\infty}^{\bar{\varepsilon}_{t}^{m}}zd\Lambda\left(z\right)+\int_{\bar{\varepsilon}_{t}^{M}}^{\infty}zd\Lambda\left(z\right)-\left(1-m_{t}\right)\delta_{t}\right)+\tau m_{t}\Delta_{t-1}\\
 & = & \frac{\eta\alpha\sigma_{x}}{1+\alpha}\left(\varphi\left(\bar{\varepsilon}_{t}^{m}\right)+\int_{-\infty}^{\infty}zd\Lambda\left(z\right)-\varphi\left(\bar{\varepsilon}_{t}^{M}\right)-\left(1-m_{t}\right)\delta_{t}\right)+\tau m_{t}\Delta_{t-1}\\
 & = & \frac{\eta\alpha\sigma_{x}}{1+\alpha}\left(\varphi\left(\bar{\varepsilon}_{t}^{m}\right)-\varphi\left(\bar{\varepsilon}_{t}^{M}\right)-\left(1-m_{t}\right)\delta_{t}\right)+\tau m_{t}\Delta_{t-1}
\end{eqnarray*}
where the second equality follows by the definition of $q_{it}^{*}$
in (\ref{eq:q_it}), and the last line follows by $\int_{-\infty}^{\infty}zd\Lambda\left(z\right)=0$,
the symmetry of $\Lambda$.

\section{Verification of Technical Results\label{sec:Verification}}

\textbf{In Section \ref{sec:Information-Based-Structural-Mod}} we
have claimed that if $\Lambda$ is unimodal, then $m_{t}$ is strictly
decreasing in $\left|\delta_{t}\right|\in\left(0,\infty\right)$.
Here we verify this claim. When $\delta_{t}>0$, by the Leibniz integral
rule
\begin{align*}
\frac{\partial m_{t}}{\partial\delta_{t}} & =\frac{\partial}{\partial\delta_{t}}\left[\Lambda\left(\delta_{t}+\zeta_{t-1}\right)-\Lambda\left(\delta_{t}-\zeta_{t-1}\right)\right]=\lambda\left(\delta_{t}+\zeta_{t-1}\right)-\lambda\left(\delta_{t}-\zeta_{t-1}\right)\\
 & =\int_{\delta_{t}-\zeta_{t-1}}^{\delta_{t}+\zeta_{t-1}}\frac{\partial\lambda\left(x\right)}{\partial x}dx=\int_{\delta_{t}}^{\delta_{t}+\zeta_{t-1}}+\int_{\delta_{t}-\zeta_{t-1}}^{\delta_{t}}\frac{\partial\lambda\left(x\right)}{\partial x}dx,
\end{align*}
where $\lambda\left(x\right)=\partial\Lambda\left(x\right)/\partial x$
is the probability density of $\Lambda$, and we assume $\lambda\left(x\right)$
is differentiable. Since $\lambda$ is symmetric and unimodal, we
have $\frac{\partial\lambda\left(x\right)}{\partial x}\big|_{x=y}+\frac{\partial\lambda\left(x\right)}{\partial x}\big|_{x=-y}=0$
for $y\in\mathbb{R}$ and $\frac{\partial\lambda\left(x\right)}{\partial x}\big|_{x=y}\leq0$
for $y\in\left(0,\infty\right)$. Given a fixed $\zeta_{t-1}$, if
$\delta_{t}\in\left(0,\zeta_{t-1}\right)$ we have 
\[
\frac{\partial m_{t}}{\partial\delta_{t}}=\int_{\zeta_{t-1}-\delta_{t}}^{\delta_{t}+\zeta_{t-1}}+\int_{0}^{\zeta_{t-1}-\delta_{t}}+\int_{\delta_{t}-\zeta_{t-1}}^{0}\frac{\partial\lambda\left(x\right)}{\partial x}dx=\int_{\zeta_{t-1}-\delta_{t}}^{\delta_{t}+\zeta_{t-1}}\frac{\partial\lambda\left(x\right)}{\partial x}dx\leq0;
\]
and obviously, $\partial m_{t}/\partial\delta_{t}\leq0$ for $\delta_{t}\in[\zeta_{t-1},\infty)$.
Parallel analysis applies when $\delta_{t}<0$.

\textbf{In Section \ref{subsec:thin-set-Identification}} we have
claimed that under the event $G_{2}$ we have $R_{t}\left(\theta\right)=\psi\left(\sqrt{\tau}\frac{1+\alpha}{\alpha\sigma_{x}\sqrt{\eta}}\left|\Delta_{t-1}\right|\right)\tau\Delta_{t-1}$.
Here we verify this claim. The event $G_{2}$ implies $\delta_{t}=0$,
under which we have $\varphi\left(\bar{\varepsilon}_{t}^{m}\right)-\varphi\left(\bar{\varepsilon}_{t}^{M}\right)=\varphi\left(-\zeta_{t-1}\right)-\varphi\left(\zeta_{t-1}\right)=0$
since for any $a\geq0$, 
\[
\varphi\left(a\right)=\int_{-\infty}^{a}zd\Lambda\left(z\right)=\int_{-\infty}^{-a}+\int_{-a}^{a}zd\Lambda\left(z\right)=\varphi\left(-a\right)+\int_{-a}^{a}zd\Lambda\left(z\right)=\varphi\left(-a\right)
\]
by the symmetry of the density of $\Lambda$ around 0. The symmetry
also implies $m_{t}=\Lambda\left(\zeta_{t-1}\right)-\Lambda\left(-\zeta_{t-1}\right)=\psi\left(\zeta_{t-1}\right)$.
Thus $R_{t}\left(\theta\right)$ in (\ref{eq_Return}) is reduced
to 
\[
R_{t}\left(\theta\right)=\tau m_{t}\Delta_{t-1}=\psi\left(\zeta_{t-1}\right)\tau\Delta_{t-1}
\]
 given $\rho=1$ and $\delta_{t}=0$.

\textbf{In Footnote \ref{fn:wG2}} we have claimed that not knowing
$\alpha$ in $W_{t}^{G_{2}}\left(\alpha,h_{T}\right)$ has no asymptotic
effect. Given the definition of $J\left(\theta\right)$ with $\alpha$
in $w_{t}^{G_{2}}\left(\alpha,h_{T}\right)$, under the regularity
conditions we have $J\left(\theta_{0}\right)\stackrel{d}{\to}\chi^{2}\left(8\right)$.
Now we consider the value of the criterion function evaluated any
$\tilde{\theta}$ on the boundary of a $T^{-1/2}$-neighborhood of
$\theta_{0}$ so that $\Vert\tilde{\theta}-\theta_{0}\Vert=cT^{-1/2}$,
where $c>0$ is some constant and $\left\Vert \cdot\right\Vert $
is the $L_{2}$-norm. $\tilde{\theta}$ is a sequence of points on
the parameter space that converges to $\theta_{0}$ at rate $T^{-1/2}$.
A Taylor expansion of $\bar{\mathbf{g}}(\tilde{\theta})$ around $\bar{\mathbf{g}}\left(\theta_{0}\right)$
gives
\begin{align*}
J(\tilde{\theta}) & =T\left(\bar{\mathbf{g}}\left(\theta_{0}\right)+\frac{\partial}{\partial\theta'}\bar{\mathbf{g}}(\check{\theta})(\tilde{\theta}-\theta_{0})\right)'\widehat{\Omega}^{-1}(\tilde{\theta})\left(\bar{\mathbf{g}}\left(\theta_{0}\right)+\frac{\partial}{\partial\theta'}\bar{\mathbf{g}}(\check{\theta})(\tilde{\theta}-\theta_{0})\right)=J\left(\theta_{0}\right)+\upsilon(\theta_{0},\check{\theta},\tilde{\theta})
\end{align*}
where $\check{\theta}$ lies on the line segment connecting $\tilde{\theta}$
and $\theta_{0}$, and 
\begin{align*}
\upsilon(\theta_{0},\check{\theta},\tilde{\theta}) & =\kappa(\theta_{0},\check{\theta},\tilde{\theta})
-2T\bar{\mathbf{g}}(\theta_{0})'\widehat{\Omega}^{-1}(\tilde{\theta})\frac{\partial}{\partial\theta^{\prime}}\bar{\mathbf{g}}(\check{\theta})(\tilde{\theta}-\theta_{0})\geq\kappa(\theta_{0},\check{\theta},\tilde{\theta})-2J^{1/2}(\theta_{0},\tilde{\theta})\kappa^{1/2}(\theta_{0},\check{\theta},\tilde{\theta})
\end{align*}
where the inequality follows the Cauchy-Schwarz inequality, and 
\begin{align*}
\kappa(\theta_{0},\check{\theta},\tilde{\theta}) & =T(\tilde{\theta}-\theta_{0})'\widehat{\Sigma}(\check{\theta},\tilde{\theta})(\tilde{\theta}-\theta_{0})\\
\widehat{\Sigma}(\check{\theta},\tilde{\theta}) & =\frac{\partial}{\partial\theta}\bar{\mathbf{g}}(\check{\theta})'\widehat{\Omega}^{-1}(\tilde{\theta})\frac{\partial}{\partial\theta'}\bar{\mathbf{g}}(\check{\theta})\\
J\left(\theta_{0},\tilde{\theta}\right) & =T\bar{\mathbf{g}}(\theta_{0})'\widehat{\Omega}^{-1}(\tilde{\theta})\bar{\mathbf{g}}(\theta_{0}).
\end{align*}

Since the non-random sequence $\tilde{\theta}\to\theta_{0}$, we have
$J(\theta_{0},\tilde{\theta})=J\left(\theta_{0}\right)+o_{p}\left(1\right)$.
On the other hand, 
\begin{align*}
\kappa(\theta_{0},\check{\theta},\tilde{\theta}) & \geq\phi_{\min}\left(\widehat{\Sigma}(\check{\theta},\tilde{\theta})\right)T\Vert\tilde{\theta}-\theta_{0}\Vert^{2}=c\cdot\phi_{\min}\left(\widehat{\Sigma}(\check{\theta},\tilde{\theta})\right)
\end{align*}
where $\phi_{\min}\left(\cdot\right)$ is the minimal eigenvalue of
a matrix. Assume $\Pr\left(\phi_{\min}\left(\widehat{\Sigma}\left(\theta_{0},\theta_{0}\right)\right)>\underline{\phi}\right)\to1$
for some constant $\underline{\phi}$ bounded away from 0, and then
we have $\kappa(\theta_{0},\check{\theta},\tilde{\theta})\geq\underline{\phi}c-o_{p}\left(1\right)$
with probability approaching one as $T\to\infty$. For any fixed constant
$c>0$, we have 
\[
\liminf_{T\to\infty}\Pr\left(4J\left(\theta_{0},\tilde{\theta}\right)<\kappa(\theta_{0},\check{\theta},\tilde{\theta})\right)>0.
\]
 When $4J\left(\theta_{0},\tilde{\theta}\right)<\kappa(\theta_{0},\check{\theta},\tilde{\theta})$
occurs, we have $\upsilon(\theta_{0},\check{\theta},\tilde{\theta})>0$
and $J(\tilde{\theta})>J\left(\theta_{0}\right)$. This argument rules
out the possibility that $\widehat{\theta}_{\mathrm{XMM}}$ is asymptotic
biased because, as the global minimizer of $J\left(\theta\right)$,
it cannot ``live'' on or outside of a neighborhood shrinking to
$\theta_{0}$ at rate $T^{-1/2}$; otherwise there is always positive
probability that $\widehat{\theta}_{\mathrm{XMM}}$ violates the definition
as an minimizer. Therefore, the effect of not knowing $\alpha$ in
$w^{G_{2}}\left(\alpha,h_{T}\right)$ does not cause asymptotic bias.
Once we have the rate of convergence, the asymptotic normality follows
from the standard XMM. 

This favorable result is driven by the one-step estimation, in which
$\widehat{\theta}_{\mathrm{XMM}}$'s convergence is guaranteed by
all the eight moments together in comparison to the ideal $J\left(\theta_{0}\right)$
that is immune from the unknown $\alpha$ in $w^{G_{2}}\left(\alpha,h_{T}\right)$.
In contrast, we do not ``plug in'' a first-step estimator of $\widehat{\alpha}^{\left(1\right)}$
into $w^{G_{2}}\left(\alpha,h_{T}\right)$ and proceed with a two-step
estimator $\widehat{\theta}^{\left(2\right)}$, where the superscript
$\left(1\right)$ and $\left(2\right)$ refer to the first step and
the second step. Such a two-step estimation method depends on the
property of $\widehat{\alpha}^{\left(1\right)}$, which may cause
asymptotic bias in $\widehat{\theta}^{\left(2\right)}$.


\end{appendices}

\bibliographystyle{chicago}
\bibliography{SmmHAM}

\newpage
\setcounter{section}{1}


\maketitle\renewcommand{\thesection}{S\arabic{section}} 
\renewcommand{\theequation}{S\arabic{equation}} 
\renewcommand{\thefigure}{S\arabic{figure}} 
\renewcommand{\thetable}{S\arabic{table}} 

 \begin{center}
		\Huge  Online Supplement
\end{center}

\bigskip

Due to space limitation, we prepare this Online Supplement for robustness
check, additional empirical results, some implementation details,
an extension of the model, and two more examples of the heterogeneous
agent model to which the technique of thin-set identification is applicable.

\section{Robustness Check: ELXM\label{subsec:Robustness-Check:-ELXM}}

Empirical likelihood \citep{qin1994empirical,kitamura1997empirical}
is an alternative to GMM. It is natural to design empirical likelihood
with extended moments (ELXM) as a counterpart of extended method of
moments (XMM). To check the robustness of the estimation results across
different methods, we estimate the model with ELXM in this section.
We first describe how to carry out ELXM. 

If the observations are i.i.d., empirical likelihood (EL) is formulated
as a constrained optimization problem 
\begin{gather*}
\max_{\theta\in\Theta,\left(\pi_{t}\in\left[0,1\right]\right)_{t=1}^{T}}\sum_{t=1}^{T}\log\pi_{t},\ \ \mbox{subject to\ \ }\sum_{t=1}^{T}\pi_{t}=1\mbox{ and }\sum_{t=1}^{T}\pi_{t}\mathbf{g}_{t}\left(\theta\right)=0,
\end{gather*}
where $\pi_{t}$ is the probability assigned to the $t$-th observation.
EL is known to be asymptotically equivalent to GMM at the first order. 

In time series, however, the blockwise EL \citep{kitamura1997empirical}
takes a distinctive scheme to account for the temporal dependence.
We propose (blockwise) ELXM for time series. Let $B_{T}$ be the block
size and $S=\left\lfloor T/B_{T}\right\rfloor $ be the number of
blocks. The blockwise moment function can be written as 
\[
g_{js}^{\left(B_{T}\right)}\left(\theta\right)=\frac{1}{B_{T}}\sum_{t=s\left(B_{T}-1\right)+1}^{sB_{T}}g_{jt}\left(\theta\right),\text{ for }s=1,\ldots,S;j=1,\ldots,8
\]
where the blockwise summation deals with time dependence. The primal
problem of ELXM is formulated as 
\begin{equation}
\max_{\theta\in\Theta,\left(\pi_{s}\in\left[0,1\right]\right)_{s=1}^{S}}\sum_{s=1}^{S}\log\pi_{s}\ \ \mbox{subject to\ \  }\sum_{s=1}^{S}\pi_{s}=1\mbox{ and }\sum_{s=1}^{S}\pi_{s}\mathbf{g}_{s}^{\left(B_{T}\right)}\left(\theta\right)=0,\label{eq:ELXM}
\end{equation}
where $\mathbf{g}_{s}^{\left(B_{T}\right)}\left(\theta\right)$ is
the blockwise counterpart of $\mathbf{g}_{t}\left(\theta\right)$,
and $\pi_{s}$ is the probability assigned to the $s$-th block. We
denote the maximizer of $\theta$ in (\ref{eq:ELXM}) as $\widehat{\theta}_{\mathrm{ELXM}}$.
ELXM is the adaption of XMM into the EL framework. Stable results
between ELXM and XMM would reinforce the robustness to the numerical
optimization procedure and the tuning parameters for time dependence. 

As an extension of XMM, \citet[pp.2109--2010]{gagliardini2011efficient}
have discussed the asymptotic distribution of XMM's EL cousin. If
$B_{T}\to\infty$, $B_{T}/T^{1/2}\to0$ as $T\to\infty$ \citep[Theorem 1(vii), p.2090]{kitamura1997empirical},
the asymptotic distribution of $\widehat{\theta}_{\mathrm{ELXM}}$
is equivalent to that of $\widehat{\theta}_{\mathrm{XMM}}$. First-order
asymptotic equivalence further indicates that the likelihood ratio
statistic
\[
LR=2\left(S\log\left(\frac{1}{S}\right)-\sum_{s=1}^{S}\log\widehat{\pi}_{s}\right)\stackrel{\mathrm{d}}{\to}\chi^{2}\left(4\right),
\]
where $\left(\widehat{\pi}_{s}\right)_{s=1}^{S}$ is the implied
probability\textemdash the maximizer of the $\left(\pi_{s}\right)_{s=1}^{S}$
part in the primal problem (\ref{eq:ELXM}). Therefore, ELXM estimator
is asymptotic equivalent to XMM, and the likelihood ratio test follows
the same asymptotic distribution as that of the $J$ test.

\begin{table}[h]
	\caption{\label{tab:tab1-EXLM}Estimation Results of ELXM for the Full Model}
	
	\bigskip{}
	\begin{centering}
		\begin{tabular}{c|cc||c|cc||c|cc||c}
			\hline 
			& \multicolumn{3}{c|}{Period 1} & \multicolumn{3}{c|}{Period 2} & \multicolumn{3}{c}{Period 3}\tabularnewline
			& est. & \multicolumn{2}{c|}{95\% CI} & est. & \multicolumn{2}{c|}{95\% CI} & est. & \multicolumn{2}{c}{95\% CI}\tabularnewline
			\hline 
			$\sigma_{\mu}$ & 0.013 & \multicolumn{2}{c|}{(0.010, 0.017)} & 0.007 & \multicolumn{2}{c|}{(0.006, 0.008)} & 0.028 & \multicolumn{2}{c}{(0.028, 0.029)}\tabularnewline
			$\eta$ & 0.106 & \multicolumn{2}{c|}{(0.087, 0.126)} & 0.167 & \multicolumn{2}{c|}{(0.125, 0.210)} & 0.218 & \multicolumn{2}{c}{(0.126, 0.310)}\tabularnewline
			$\tau$ & 0.597 & \multicolumn{2}{c|}{(0.430, 0.763)} & 0.706 & \multicolumn{2}{c|}{(0.422, 0.989)} & 0.857 & \multicolumn{2}{c}{(0.415, 1.300)}\tabularnewline
			$\alpha$ & 1.551 & \multicolumn{2}{c|}{(0.642, 2.461)} & 1.863 & \multicolumn{2}{c|}{(0.946, 2.779)} & 3.240 & \multicolumn{2}{c}{(0.808, 5.673)}\tabularnewline
			LR-stat. & \multicolumn{3}{c|}{4.205} & \multicolumn{3}{c|}{5.121 } & \multicolumn{3}{c}{8.684}\tabularnewline
			\emph{p}-value & \multicolumn{3}{c|}{(0.379)} & \multicolumn{3}{c|}{(0.275)} & \multicolumn{3}{c}{ (0.069)}\tabularnewline
			\hline 
		\end{tabular}
		\par\end{centering}
	\bigskip{}
	\raggedright{}\small Note: Similar to Table 1, the likelihood ratio
	statistics (LR-stat.) of the over-identification test follows $\chi^{2}\left(4\right)$
	asymptotic distribution under the null.
\end{table}

The numerical implementation of ELXM is similar to the standard blockwise
EL. We carry out the numerical optimization in two steps: (i) solve
$\pi$ in the inner step given a trial value $\theta$, and (ii) solve
$\theta$ in the outer step. We optimize the convex primal problem
in the inner loop, while the outer step is a standard low-dimensional
nonlinear optimization. We set $B_{T}$ equal to the number of lags
for the long-run variance calculation in XMM in Section 3.4 of the
main text.

The ELXM estimation of the full model is reported in Table \ref{tab:tab1-EXLM}.
The point estimates and the confidence intervals are close to those
of XMM. The point estimates also yield very similar predicted moments
as XMM in Table 2 and the switching between fundamentalists and chartists
exhibits similar patterns with those in Figures 1, 2, and 3, which
we do not repeat here. Nevertheless, the LR test statistics of Period
3 is 8.68, with a $p$-value of 0.07. The over-identification test
rejects the null hypothesis at 10\% significance level. The evidence
of marginal rejection echoes the big $J$-statistic for Period 3 in
Table 1 of the main text. It indicates that we must be cautious when
applying our model to a long time span with high volatility and potential
structural changes. 

For further comparison, we run the standard blockwise EL to estimate
the model with unconditional moments, as we did for GMM. The results
are displayed in Table \ref{tab:tab1-EL}. Again, we observe the pattern
of smaller $\eta$ and wider confidence intervals, which echoes that
in Table 3. We also try ELXM for the solo-strategy models, in which
we encounter the numerical problem of infeasible constraints in all
three periods. The infeasibility problem is well understood in the
literature of EL as strong evidence of model misspecification \citep{chen2008adjusted}.
Severe model misspecification is manifest in the very small $p$-values
in Table 4. The evidence from the ELXM and EL estimation suggests
robustness of the empirical results in the full model and the model
with the unconditional moments, as well as strong rejection of the
solo-strategy models.

\begin{table}[h]
	\caption{\label{tab:tab1-EL}Estimation Results of EL for the Unconditional
		Moment Model}
	
	\bigskip{}
	\begin{centering}
		\begin{tabular}{c|cc||c|cc||c|cc||c}
			\hline 
			& \multicolumn{3}{c|}{Period 1} & \multicolumn{3}{c|}{Period 2} & \multicolumn{3}{c}{Period 3}\tabularnewline
			& est. & \multicolumn{2}{c|}{95\% CI} & est. & \multicolumn{2}{c|}{95\% CI} & est. & \multicolumn{2}{c}{95\% CI}\tabularnewline
			\hline 
			$\sigma_{\mu}$ & 0.014 & \multicolumn{2}{c|}{(0.014, 0.015)} & 0.007 & \multicolumn{2}{c|}{(0.007, 0.007)} & 0.029 & \multicolumn{2}{c}{(0.029, 0.030)}\tabularnewline
			$\eta$ & 0.116 & \multicolumn{2}{c|}{(0.007, 0.225)} & 0.126 & \multicolumn{2}{c|}{(-0.218, 0.470)} & 0.060 & \multicolumn{2}{c}{(-0.004, 0.124)}\tabularnewline
			$\tau$ & 0.678 & \multicolumn{2}{c|}{(0.439, 0.917)} & 0.625 & \multicolumn{2}{c|}{(0.116, 1.133)} & 0.750 & \multicolumn{2}{c}{(0.421, 1.080)}\tabularnewline
			$\alpha$ & 2.644 & \multicolumn{2}{c|}{(0.879, 4.408)} & 1.751 & \multicolumn{2}{c|}{(-0.135, 3.636)} & 3.672 & \multicolumn{2}{c}{(2.200, 5.145)}\tabularnewline
			LR-stat. & \multicolumn{3}{c|}{0.037 } & \multicolumn{3}{c|}{1.103} & \multicolumn{3}{c}{0.234}\tabularnewline
			\emph{p}-value & \multicolumn{3}{c|}{(0.848)} & \multicolumn{3}{c|}{(0.294)} & \multicolumn{3}{c}{ (0.628)}\tabularnewline
			\hline 
		\end{tabular}\bigskip{}
		\par\end{centering}
	\raggedright{}\small Note: Similar to Table 3, the likelihood ratio
	statistic of the over-identification test follows $\chi^{2}\left(1\right)$
	asymptotic distribution under the null.
\end{table}

\section{Additional Empirical Results}

In the main text, valid inference relies on several assumptions in
the structural model. This section presents additional empirical results
to verify some assumptions. 

\textbf{Local identification.} In the main text we have assumed local
identification, following \citet{gagliardini2011efficient} and \citet{antoine2012efficient}.
Here we provide statistical evidence of local identification. Local
identification is equivalent to a full-rank Jacobian matrix. We use
\citet{kleibergen2006generalized}'s reduced-rank test (KP test) to
check the rank of the empirical Jacobian matrix $\widehat{H}_{\mathrm{unc}}\left(\theta\right)=\frac{\partial}{\partial\theta'}\overline{\mathbf{g}}_{\mathrm{unc}}\left(\theta\right)$,
where $\overline{\mathbf{g}}_{\mathrm{unc}}=\left(\overline{\mathbf{g}}_{j}\right)_{j\in\left\{ 1,5,\ldots,8\right\} }$
is the vector of the 5 unconditional sample moments. The data support
a full rank $\widehat{H}_{\mathrm{unc}}\left(\theta\right)$ if we
can reject the null hypothesis that its rank is 3, 2, or 1. We evaluate
the rank of $\widehat{H}_{\mathrm{unc}}\left(\theta\right)$ at either
$\widehat{\theta}_{\mathrm{XMM}}$ or $\widehat{\theta}_{\mathrm{GMM}}$.
Table \ref{tab:KP-test} reports the KP test statistics under the
null of rank 3. We have also conducted the same test under the null
that the rank of $\widehat{H}_{\mathrm{unc}}\left(\theta\right)$
is 2 or 1, respectively, and the rejection is overwhelming in all
cases. The KP test provides evidence of non-trivial local information
from the unconditional moments. 

\begin{table}
	\caption{\label{tab:KP-test}KP Test Statistic and $p$-value}
	
	\bigskip{}
	\begin{centering}
		\begin{tabular}{c|c||c||c|c||c||c|c||c||c}
			\hline 
			& \multicolumn{3}{c|}{Period 1} & \multicolumn{3}{c|}{Period 2} & \multicolumn{3}{c}{Period 3}\tabularnewline
			XMM & \multicolumn{3}{c|}{15.850 (0.000)} & \multicolumn{3}{c|}{5.826 (0.054)} & \multicolumn{3}{c}{7.319 (0.026)}\tabularnewline
			GMM & \multicolumn{3}{c|}{19.141 (0.000)} & \multicolumn{3}{c|}{7.251 (0.027)} & \multicolumn{3}{c}{9.563 (0.008)}\tabularnewline
			\hline 
		\end{tabular}
		\par\end{centering}
	\bigskip{}
	\raggedright{}\small Note: The null hypothesis is that the rank of
	$\widehat{H}_{\mathrm{unc}}\left(\theta\right)$ is 3. The $p$-value
	in the parenthesis is calculated according to the asymptotic distribution
	$\chi^{2}\left(2\right)$. 
\end{table}

\textbf{Unit root test for }$\boldsymbol{\mu}^{T}$\textbf{.} The
1\%, 5\%, and 10\% critical value for the standard Dicky-Fuller test
are -2.58, -1.95, -1.62 respectively. This is a one-sided test that
rejects the null of unit root behavior if the test statistic is smaller
than the critical value. We run the Dicky-Fuller test, and obtain
the test statistics 0.0702, 0.9672, and -0.4081 for Period 1, 2 and
3, respectively. These statistics are in favor of the null hypothesis
of the unit root. Formally, they do not reject the null of unit root
at 10\% significance level, since none of the statistics are smaller
than $-1.62$. What is more, the positive statistics in period 1 and
2, which are associated with autoregressive coefficient estimates
of 1.0004 and 1.0024, respectively, may indicate possibly very weak
explosive behavior.

\begin{table}
	\caption{\label{tab:correlation}Correlation Coefficients of the Real and Predicted
		returns}
	
	\bigskip{}
	\begin{centering}
		\begin{tabular}{llcrrr}
			\hline 
			&  & \multirow{2}{*}{Real} & \multicolumn{1}{c}{XMM} & \multirow{2}{*}{GMM} & \multicolumn{1}{c}{XMM }\tabularnewline
			&  &  & \multicolumn{1}{c}{full} &  & fund.-only\tabularnewline
			\hline 
			& XMM full model & 0.047 &  &  & \tabularnewline
			Period 1 & GMM  & 0.048 & 0.975 &  & \tabularnewline
			& XMM fundamentalist-only & 0.075 & 0.613 & 0.565 & \tabularnewline
			& XMM chartist-only & 0.165 & 0.279 & 0.281 & -0.197\tabularnewline
			\hline 
			& XMM full model & 0.117 &  &  & \tabularnewline
			Period 2 & GMM & 0.127 & 0.949 &  & \tabularnewline
			& XMM fundamentalist-only & 0.092 & 0.582 & 0.484 & \tabularnewline
			& XMM chartist-only & 0.081 & 0.201 & 0.313 & -0.374\tabularnewline
			\hline 
			& XMM full model & 0.112 &  &  & \tabularnewline
			Period 3 & GMM  & 0.130 & 0.821 &  & \tabularnewline
			& XMM fundamentalist-only & 0.014 & 0.410 & 0.077 & \tabularnewline
			& XMM chartist-only & 0.151 & 0.173 & 0.473 & -0.538\tabularnewline
			\hline 
		\end{tabular}
		\par\end{centering}
	\bigskip{}
	\raggedright{}\small  Note: the time series here are the same as
	those in Table 2. The the entries are pairwise correlation coefficient. 
\end{table}

\textbf{Correlation.} Table \ref{tab:correlation} reports the pairwise
correlation coefficients of the real return time series $\left(\boldsymbol{R}_{t}^{\mathrm{r}}\right)$
and the predicted $\left(R_{t}\left(\theta\right)\right)$ evaluated
at the various estimates. The entries of the first column of Table
\ref{tab:correlation} are small, indicating weak correlation between
the real return and the predicted ones. This is not surprising since
we fit the moments of the marginal distribution of the returns, rather
than the temporal co-movements, to estimate the parameters. %

\begin{figure}
	\begin{centering}
		\includegraphics[scale=0.65]{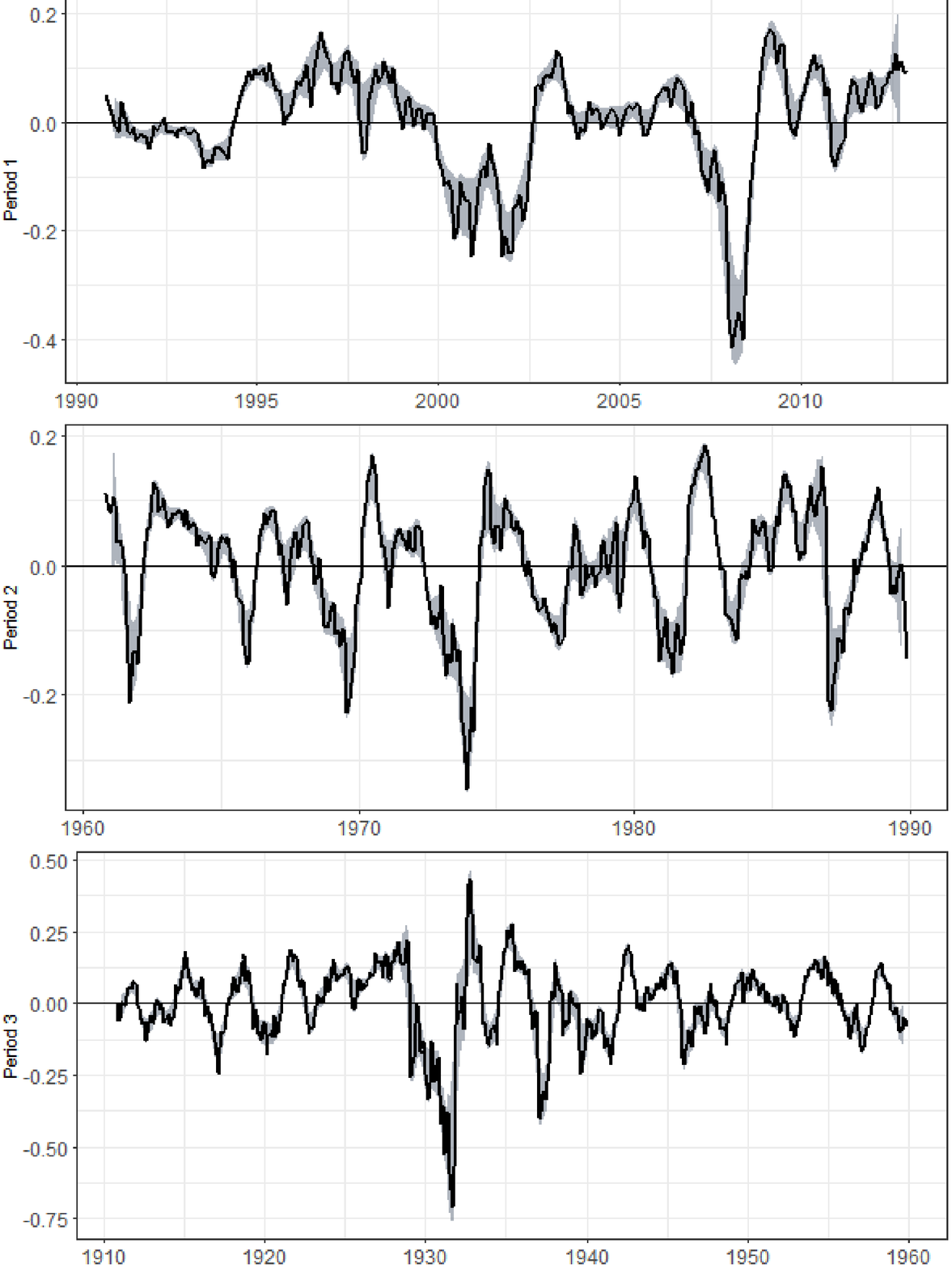}
		\par\end{centering}
	\small
	
	Note: The gray shaded region is the 90\% pointwise confidence interval
	constructed by the time series kernel smoothing method \citep[p.218]{fan2003book}.
	We use the Bartlett kernel with the same bandwidth as in the main
	text.
	
	\caption{\label{fig:Delta} $\left(\Delta_{t}\right)_{t=1}^{T}$ in All the
		Three Periods}
\end{figure}

\textbf{Frequency of the event $G_{1}$}. The analysis of the thin-set
identification starts from the event $G_{1}=\left\{ \Delta_{t-1}=0\right\} $,
and the convergence rate of the local moments depends on how often
$\Delta_{t-1}$ fluctuates around 0. Figure \ref{fig:Delta} plots
the series $\left(\Delta_{t}\right)_{t=1}^{T}$ in all the time periods.
We observe the curve vacillates around 0 repeatedly, so that $G_{1}$
is not a rare event. 

\begin{table}[h]
	\caption{\label{Markov}Two-regime Markov Switching Model}
	
	\bigskip{}
	\centering{}%
	\begin{tabular}{ll|rc||c|rc||c|rc||c}
		\hline 
		&  & \multicolumn{3}{c|}{Period 1} & \multicolumn{3}{c|}{Period 2} & \multicolumn{3}{c}{Period 3}\tabularnewline
		&  & est. & \multicolumn{2}{c|}{s.e.} & est. & \multicolumn{2}{c|}{s.e.} & est. & \multicolumn{2}{c}{s.e.}\tabularnewline
		\hline 
		Regime 1: Boom & Intercept & 0.183 & \multicolumn{2}{r|}{0.017} & 0.148 & \multicolumn{2}{r|}{0.008} & -0.003 & \multicolumn{2}{r}{0.009}\tabularnewline
		& Slope & -0.481 & \multicolumn{2}{r|}{0.180} & 0.679 & \multicolumn{2}{r|}{0.088} & 1.864 & \multicolumn{2}{r}{0.033}\tabularnewline
		Regime 2: Bust & Intercept & -0.276 & \multicolumn{2}{r|}{0.010} & -0.198 & \multicolumn{2}{r|}{0.011} & -0.132 & \multicolumn{2}{r}{0.013}\tabularnewline
		& Slope & -0.824 & \multicolumn{2}{r|}{0.101} & 0.348 & \multicolumn{2}{r|}{0.123} & 0.446 & \multicolumn{2}{r}{0.038}\tabularnewline
		Transition  & Boom$\rightarrow$Bust & \multicolumn{3}{c|}{0.018 } & \multicolumn{3}{c|}{0.020} & \multicolumn{3}{c}{ 0.025}\tabularnewline
		Probability & Bust$\rightarrow$Boom & \multicolumn{3}{c|}{0.019} & \multicolumn{3}{c|}{0.015} & \multicolumn{3}{c}{0.021}\tabularnewline
		\hline 
	\end{tabular}
\end{table}

\begin{figure}
	\begin{centering}
		\includegraphics[scale=0.65]{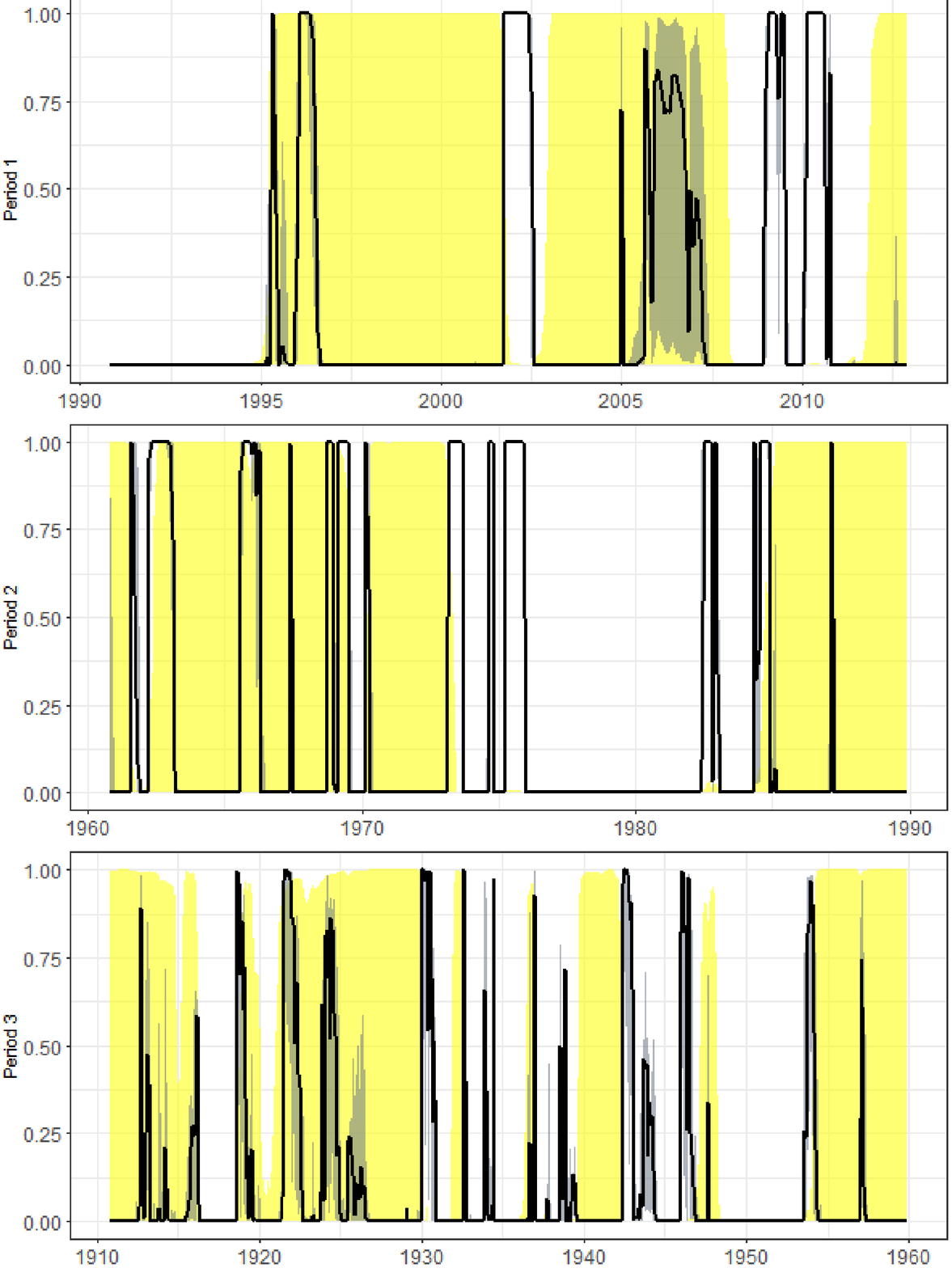}
		\par\end{centering}
	\small
	
	Note: The gray shaded region is the 90\% pointwise confidence interval
	constructed by parametric bootstrap. The yellow shaded region is the
	probability of the boom regime estimated from the two-regime Markov
	switching model.
	
	\caption{\label{fig:CI} Fraction of Chartists and Markov Switching}
\end{figure}

\textbf{Markov Switching.} We conduct a simple Markov switching model
in which we allow two regimes for the intercept and the slope coefficient
in the regression $R_{t}^{\mathrm{r}}=\mathrm{intercept}+\mathrm{slope}\times\mu_{t}+\mathrm{error\ term}$.
We refer to the regime with greater intercept and slope coefficient
as the \emph{boom regime} and the other as the \emph{bust regime}.
The boom regimes for sample period 1, 2 and 3 are shaded in yellow
color in the upper, middle and bottom panel of Figure \ref{fig:CI},
respectively. 

The boom (bust) regimes correspond to the scenarios when the market
price is rising (falling). The probability for the market to transit
from a boom to a bust ranges from 1.8\% to 2.5\%, which implies that
on average it takes 40 to 56 months for the price to reverse its trend.
Similarly, the probability for the market to transit from a bust to
a boom is very low.

There are considerable overlap between the boom regimes identified
by the Markov switching model and the chartists-dominated regime uncovered
from the structural model. It suggests that our model based on the
dynamic transition in the market fraction of chartists reasonably
captures the price movement. 

\textbf{Confidence interval of $m_{t}\left(\widehat{\theta}_{\mathrm{XMM}}\right)$.
}The fraction of the chartist, $m_{t}\left(\theta_{0}\right)$, is
a nonlinear function of $\mathbf{p}^{T}$ and $\boldsymbol{\mu}^{T}$.
In principle we can construct the pointwise confidence interval by
the delta method based on the asymptotic distribution of $\widehat{\theta}_{\mathrm{XMM}}$.
However, it is difficult to interpret when the two-sided symmetric
confidence interval goes beyond $\left[0,1\right]$, which occurred
in our experiment. 

To avoid such difficulty, we can use the parametric bootstrap if we
are willing to impose the normality assumption $\varepsilon_{t}^{\mu}\sim\mathrm{i.i.d.}N\left(0,1\right)$.
Let $\widehat{\theta}_{\mathrm{XMM}}^{*\left(b\right)}$ be a bootstrap
estimator where the superscript ``$\left(b\right)$'' indexes the
instance of bootstrap replication, and $m_{t}^{*\left(b\right)}=m_{t}\left(\widehat{\theta}_{\mathrm{XMM}}^{*\left(b\right)}\right)$
is the plug-in bootstrap estimator of the fraction. The parametric
bootstrap is implemented as follows. We simulate a sequence $\boldsymbol{\mu}^{T*\left(b\right)}=\left(\mu_{t}^{*\left(b\right)}\right)_{t=1}^{T}$
where $\mu_{t}^{*\left(b\right)}=\mu_{t-1}+\widehat{\sigma}_{\mu,\mathrm{XMM}}\varepsilon_{t}^{\mu*\left(b\right)}$
with $\varepsilon_{t}^{\mu*\left(b\right)}\sim\mathrm{i.i.d.}N\left(0,1\right)$.
Given the data $\left(\boldsymbol{\mu}^{T*\left(b\right)},\mathbf{p}^{T}\right)$,
we obtain the bootstrap estimator $\widehat{\theta}_{\mathrm{XMM}}^{*\left(b\right)}$.
Here we only bootstrap $\boldsymbol{\mu}^{T*\left(b\right)}$ since
the function $R_{t}\left(\theta\right)$ only depends on the $\left(p_{t-1},p_{t-1}^{c},\mu_{t},\mu_{t-1}\right)$
but not $p_{t}$. After having $\widehat{\theta}_{\mathrm{XMM}}^{*\left(b\right)}$,
we plug it into $m_{t}\left(\theta\right)$ and get $m_{t}^{*\left(b\right)}$.
We repeat the bootstrap for 199 times, and compute the 5\% and 95\%
sample quantiles of $\left(m_{t}^{*\left(b\right)}\right)_{b=1}^{200}$
for each $t$ as the lower and upper bounds of the 90\% two-sided
pointwise confidence interval for $m_{t}\left(\theta_{0}\right)$.

The estimated pointwise bootstrap confidence interval is shown as
the gray shaded region in Figure \ref{fig:CI}. The confidence interval
is very narrow most of the time, in particular when $m_{t}\left(\widehat{\theta}_{\mathrm{XMM}}\right)$
is close to 0. Interestingly, along with the swings of the fraction
of the chartists before the 2008 financial crisis, the uncertainty
is manifest by the relatively wide confidence intervals.

\section{Implementation}

\subsection{Gordon Growth Model\label{subsec:Gordon}}

The original Gordon growth model is defined as $\tilde{\mu}=d_{t}(1+\kappa)/(\beta-\kappa),$
where $d_{t}$ is the dividend at period $t$, $\beta$ is the discount
rate and $\kappa$ is the average growth rate of dividends. \citet{fama2002equity}
suggest that the Gordon growth model implies $\beta=\bar{y}+\kappa$,
where $\bar{y}$ is the average dividend yield. We replace $\beta$
by $\bar{y}+\kappa$ and obtain $\mu_{t}=d_{t}(1+\kappa)/\bar{y}.$

\subsection{Chartist-Only Model\label{subsec:The-Chartist-Only-Model}}

Unlike the fundamentalist-only model, the chartist-only model is not
a sub-model of the benchmark model, since $\eta$ cannot be set as
0. Even if we treat $\tau/0=\infty$, or view the model as a sequence
of models with $\eta\to0^{+}$, the three kernel-weighted moment functions
still break down. When $\eta$ becomes arbitrarily small, the fundamental
strategy will return infinitesimal profit. It violates the assumption
that the fundamental strategy beats the chartist strategy under arbitrarily
deviation from $\Delta_{t-1}=0$, and invalidates $g_{2t}\left(\theta\right)$
and $g_{3t}\left(\theta\right)$, which were justified by arguing
that the market is dominated by fundamentalists when $G_{1}$ occurs.
Moreover, as a chartist ignores the fundamental value, $\alpha$ is
also unidentified; thus $g_{4t}\left(\theta\right)$ is not well defined.

Given the difficulty of adapting it to the chartist-only scenario,
we slightly modify the benchmark model. In a market with only chartists,
the demand equation becomes $R_{t}\left(\theta\right)=\tau\Delta_{t-1}.$
Notice that even without fundamentalists, the event $G_{1}$ remains
well defined. Thus we introduce another kernel-weighted moment function
\[
g_{9t}\left(\theta\right)=w_{t}^{G_{1}}\left(h_{T}\right)\left(\left|R_{t}^{\mathrm{r}}\right|-\tau\left|\Delta_{t-1}\right|\right),
\]
where $G_{2}$ is replaced by $G_{1}$. This conditional moment is
implied by the chartist-only model: when $\Delta_{t-1}$ is close
to 0, the return must also be small.

When estimating the chartist-only model, we utilize $g_{9t}\left(\theta\right)$
along with the five unconditional moment functions $\left\{ g_{jt}\left(\theta\right)\right\} _{j=1,5,6,7,8}$.
The estimation involves six moments and two parameters $\left(\sigma_{\mu},\tau\right)$,
so that the $J$-statistic still follows $\chi^{2}\left(4\right)$
asymptotically under the null hypothesis.Extension of the Model

\section{Extension of the Model}

Homogeneity and time invariance of the conditional variance in returns
is restrictive, especially during a financial crisis. In this section,
we discuss the possibility of relaxing this assumption. 

We define $\eta_{it}=1/\left(A^{f}\cdot\mbox{var}{}_{it-1}^{f}[R_{t}^{f}]\right)$
to allow $\mbox{var}{}_{it-1}^{f}[R_{t}^{f}]$ to vary across $i$
and $t$. Similarly, define $\tau_{t}=1/\left(A^{c}\cdot\mbox{var}{}_{t-1}^{c}\left[R_{t}^{c}\right]\right)$,
which is time-varying but individual invariant as the chartist strategy
does not consider any private signal. It follows that for the fundamental
strategy $\pi_{it}^{f}=\eta_{it}\left(\frac{\alpha\sigma_{x}}{1+\alpha}\left(\varepsilon_{it}-\delta_{t}\right)\right)^{2}$,
and for the chartist strategy $\pi_{t}^{c}=\tau_{t}\Delta_{t-1}^{2}$.
The investor chooses the fundamental strategy if $\pi_{it}^{f}\geq\pi_{t}^{c}$,
and the demand of the risky asset is 
\[
q_{it}^{*}=q_{it}^{f*}\cdot\boldsymbol{1}\left\{ \pi_{it}^{f}\geq\pi_{t}^{c}\right\} +q_{it}^{f*}\cdot\boldsymbol{1}\left\{ \pi_{it}^{f}<\pi_{t}^{c}\right\} .
\]
To compute the market aggregate demand, we need to specify the conditional
variances since neither $\mbox{var}{}_{it-1}^{f}[R_{t}^{f}]$ nor
$\mbox{var}{}_{t-1}^{c}\left[R_{t}^{c}\right]$ is observable from
the data. A simple rule from the observed past history is an option
for the chartist, while there is no consensus in the literature about
the conditional variance of the fundamental strategy. 

Consider imposing a parametric assumption on the joint distribution
of $\left(\eta_{it},\varepsilon_{it}\right)$, for example, jointly
normal i.i.d.~across time. This simple specification introduces two
extra parameters: the variance of $\eta_{it}$ that captures the dispersion
of the beliefs on the volatility, and the correlation coefficient
between $\eta_{it}$ and $\varepsilon_{it}$. Although the theoretical
model can be simulated by the method of simulated moments (MSM), all
the closed-forms in the aggregate demand and the thin-set identification
are lost. Such difficulty arises even before we study any dynamic
specification in $\left(\eta_{it}\right)$, which will incur additional
parameters. 

Analysis becomes more tractable if we assume that the distribution
of $\eta_{it}$ and $\varepsilon_{it}$ are independent.\textbf{ }Let
$\bar{\varepsilon}_{it}$ be the threshold such that $\pi_{it}^{f}=\pi_{t}^{c}$.
We solve $\eta_{it}\left(\frac{\alpha\sigma_{x}}{1+\alpha}\left(\bar{\varepsilon}_{it}-\delta_{t}\right)\right)^{2}=\tau_{t}\Delta_{t-1}^{2}$
to obtain 
\[
\bar{\varepsilon}_{it}=\delta_{t}\pm\frac{1+\alpha}{\alpha\sigma_{x}}\sqrt{\frac{\tau_{t}}{\eta_{it}}}\left|\Delta_{t-1}\right|=\delta_{t}\pm\zeta_{it-1},
\]
where $\zeta_{it-1}=\frac{1+\alpha}{\alpha\sigma_{x}}\sqrt{\frac{\tau_{t}}{\eta_{it}}}\left|\Delta_{t-1}\right|$.
Define the lower bound $\bar{\varepsilon}_{it}^{m}=\delta_{t}-\zeta_{it-1}$
and upper bound $\bar{\varepsilon}_{it}^{M}=\delta_{t}+\zeta_{it-1}$.
As a result, the individual investment flow is 
\[
q_{it}^{*}=q_{it}^{f*}\cdot\mathbf{1}\left\{ \varepsilon_{it}\in(-\infty,\bar{\varepsilon}_{it}^{m}]\cup[\bar{\varepsilon}_{it}^{M},\infty)\right\} +q_{t}^{c*}\cdot\mathbf{1}\left\{ \varepsilon_{it}\in(\bar{\varepsilon}_{it}^{m},\bar{\varepsilon}_{it}^{M})\right\} .
\]
The probability of individual $i$ adopting the chartist strategy
is $m_{it}=\Lambda\left(\bar{\varepsilon}_{it}^{M}\right)-\Lambda\left(\bar{\varepsilon}_{it}^{m}\right)$,
and the aggregate demand in the market is 
\begin{eqnarray}
D_{t}\left(\theta\right) & = & \int_{0}^{1}\int_{(-\infty,\bar{\varepsilon}_{it}^{m}]\cup[\bar{\varepsilon}_{it}^{M},\infty)}\frac{\eta_{it}\alpha\sigma_{x}}{1+\alpha}\left(\varepsilon_{it}-\delta_{t}\right)d\Lambda\left(\varepsilon_{it}\right)di+\tau_{t}\Delta_{t-1}\int_{0}^{1}m_{it}di\nonumber \\
& = & \frac{\alpha\sigma_{x}}{1+\alpha}\left(\int_{0}^{1}\eta_{it}\left[\int_{-\infty}^{\bar{\varepsilon}_{it}^{m}}zd\Lambda\left(z\right)+\int_{\bar{\varepsilon}_{it}^{M}}^{\infty}zd\Lambda\left(z\right)-\left(1-m_{it}\right)\delta_{t}\right]di\right)+\tau_{t}\Delta_{t-1}\int_{0}^{1}m_{it}di\nonumber \\
& = & \frac{\alpha\sigma_{x}}{1+\alpha}\left(\int_{0}^{1}\eta_{it}\left[\varphi\left(\bar{\varepsilon}_{it}^{m}\right)-\varphi\left(\bar{\varepsilon}_{it}^{M}\right)-\left(1-m_{it}\right)\delta_{t}\right]di\right)+\tau_{t}\Delta_{t-1}\int_{0}^{1}m_{it}di.\label{eq:g_demand}
\end{eqnarray}

If we further assume $\tau_{t}=\tau/\varepsilon_{t}^{c}$ with $\varepsilon_{t}^{c}$
being the proxy for $\mbox{var}{}_{t-1}^{c}\left[R_{t}^{c}\right]$,
we can pointly identify $\tau$ under the event $G_{2}$ as in the
main text. This identified $\tau$ will depend on the choice of $\varepsilon_{t}^{c}$.
On the other hand, if we assume $\eta_{it}=\eta/\varepsilon_{it}^{v}$,
where $\varepsilon_{it}^{v}$ is the shock to each individual's conditional
variance independent of all other random variables, then the identification
of $\left(\eta,\alpha\right)$ remains under the event $G_{1}$. Therefore,
in this generalized model in which we allow time-varying and heterogeneous
$\eta_{it}$, we are able to pointly identify the same parameter $\left(\eta,\alpha\right)$
as in the main text where a constant $\mbox{var}{}_{t-1}^{f}\left[R_{t}^{f}\right]$
is assumed. As a result, the empirical estimates of $\left(\eta,\alpha\right)$
in the three periods in the main text are informative about the magnitude
of these parameters.

If we start with the general model, nevertheless, MSM will be necessary
to handle the integrals $\int_{0}^{1}\eta_{it}\left[\varphi\left(\bar{\varepsilon}_{it}^{m}\right)-\varphi\left(\bar{\varepsilon}_{it}^{M}\right)-\left(1-m_{it}\right)\delta_{t}\right]di$
and $\int_{0}^{1}m_{it}di$ in the demand equation (\ref{eq:g_demand}).
We do not have simple closed-forms for these integrals as $\overline{\varepsilon}_{it}^{m}$,
$\overline{\varepsilon}_{it}^{M}$ and $m_{it}$ all depend on $\eta_{it}$
and $\tau_{t}$. Exploration of the conditional variance in this heterogeneous
agent model would contribute to the theoretical modeling, asymptotic
property of XMM-MSM, as well as the empirical findings. All these
three aspects are new to the existing literature and they deserve
thorough investigation in future research. 

\section{Examples of Thin-Set Identification \label{sec:More-Examples}}

Thin-set identification is not peculiar to our model. It is also useful
for other heterogeneous agent models. Here we give two examples.
\begin{example}
	\citet{lux1995herd} formalizes herd behavior in speculative markets
	in which bubbles emerge as self-organizing process of infection among
	traders. Let $x$ be an index ranging from $-1$ (extremely pessimistic)
	to $1$ (extremely optimistic). It characterizes the average opinion
	of speculative investors. The dynamics of $x$ is governed by the
	differential equation 
	\[
	dx/dt=2v\left(\tanh\left(ax\right)-x\cosh\left(ax\right)\right),
	\]
	where $a$ is a measure of the strength of herd behavior, and $v$
	is a variable for the speed of change. The fraction of optimistic
	trader is $0.5\left(x+1\right)\in\left[0,1\right]$ \citep[pp.884--885]{lux1995herd}.
	When $x=1$, the fraction of optimistic trader is 1.\qed
\end{example}
\begin{example}
	\citet{he2005commodity} analyze the creation of bull or bear market
	via nonlinear interactions between market participants\textemdash consumers,
	producers and heterogeneous speculators\textemdash in a behavioral
	commodity market model. They model the market share of chartists as
	$1/\left(1+d\left(F-S_{t}\right)^{2}\right)$, where $d$ is a switching
	parameter, $F$ is the long-run equilibrium price, and $S_{t}$ is
	the commodity price at time $t$ \citep[p.1582]{he2005commodity}.
	When $F=S_{t}$, the fraction of chartists is 1.\qed
\end{example}

\end{document}